\documentclass[10pt,twocolumn,showpacs,amsmath,amssymb,aps,prx,longbibliography,superscriptaddress,floatfix]{revtex4-1}
\usepackage{times}
\usepackage[utf8]{inputenc}
\usepackage{color}
\usepackage{array}
\usepackage{verbatim}
\usepackage{multirow}
\usepackage{amsmath}
\usepackage{amssymb}
\usepackage{graphicx}
\usepackage{esint}
\usepackage[unicode=true,
 bookmarks=true,bookmarksnumbered=true,bookmarksopen=false,
 breaklinks=true,pdfborder={0 0 0},backref=false,colorlinks=true]
 {hyperref}
\usepackage{enumitem}
\usepackage{cleveref}
\usepackage{bm}
\usepackage{newtxtext}
\usepackage{listings}
\usepackage{xcolor}
\usepackage[bb=boondox]{mathalfa}
\usepackage[title]{appendix}
\usepackage{makecell}
\usepackage{pifont}
\usepackage{diagbox}
\usepackage{tcolorbox}
\usepackage{soul}

\hypersetup{
    linkcolor=red,
    citecolor=blue,
    filecolor=magenta,
    urlcolor=magenta
}

\usepackage{cleveref}
\crefname{appendix}{Appendix}{Appendices}
\crefname{subappendix}{Appendix}{Appendices}
\Crefname{subappendix}{Appendix}{Appendices}
\crefname{equation}{Eq.}{Eqs.}
\crefname{figure}{Fig.}{Figs.}
\crefname{table}{Table}{Tables}
\crefname{section}{Sec.}{Secs.}

\tcbuselibrary{listings,skins}
\newcommand{\matlabcode}[2][]{
    \begin{tcblisting}{
        colback=white,
        listing only,
        enhanced,
        listing options={
            language=Matlab,
            basicstyle=\ttfamily\small,
            keywordstyle=\color{blue},
            commentstyle=\color{green},
            stringstyle=\color{red},
            showstringspaces=false,
            numbers=left,
            numberstyle=\tiny\color{gray},
            breaklines=true
        },
        #1
    }
    #2
    \end{tcblisting}
}
\renewcommand{\paragraph}[1]{\vspace{0.2cm}{\bf \textit{#1}}}

\def\eg{{\it e.g.},\ }

\definecolor{Gray}{gray}{0.85}
\newcolumntype{a}{>{\columncolor{Gray}}c}

\usepackage{simpler-wick}
\allowdisplaybreaks[1]

\newcommand{\td}{\widetilde}

\newcommand{\hH}{\hat{H}}

\def\tr{\mathrm{Tr}}

\def\up{\uparrow}
\def\down{\downarrow}

\def\ii{{\rm i}}

\def\kk{ {\mathbf{k}} }

\def\tk{T_{\rm K}}

\begin{document}
\title{Universality of Energy-Space Entanglement in Quantum Impurity Models}

\author{Geng-Dong Zhou}
\affiliation{International Center for Quantum Materials, School of Physics, Peking University, Beijing 100871, China}

\author{Zhi-Da Song}
\email{songzd@pku.edu.cn}
\affiliation{International Center for Quantum Materials, School of Physics, Peking University, Beijing 100871, China}
\affiliation{Hefei National Laboratory, Hefei 230088, China}
\affiliation{Beijing Key Laboratory of Quantum Devices, Peking University, Beijing 100871, China}

\date{\today}
\begin{abstract}
    Entanglement entropy (EE) is most commonly studied using real-space bipartitions.
Here we show that, in quantum impurity models, an energy-space bipartition, which is also the momentum-space bipartition of the bath degrees of freedom, can also display universal behaviors. We logarithmically discretize the bath and partition it into high- and low-energy sectors, motivated by poor man's scaling.
For quantum impurity models with Fermi-liquid fixed points, including the Anderson model and fully screened or underscreened Kondo models, the low-energy EE flows to constants independent of model parameters.
These constants are integer multiples of $\ln 2$ plus corrections that depend only on the logarithmic discretization parameter $\Lambda$.
We further show that the scale invariance of the fixed-point wavefunction in energy space maps to an effective translation invariance of a one-dimensional chain, allowing the impurity fixed points to be classified using topological band theory in one dimension.
With low-energy chiral symmetry, each $\ln 2$ contribution to the EE then originates from a topological edge mode.
We also study transitions between distinct Fermi-liquid fixed points, focusing on the local-singlet--Kondo-singlet transition in a two-orbital Anderson model induced by an inter-orbital antiferromagnetic coupling as an example.
The local-singlet phase has an effectively decoupled impurity and nearly vanishing EE, whereas the Kondo-singlet phase has a finite EE larger than $\ln 2$ per spin and orbital.
When chiral symmetry holds at low energies, this distinction admits a topological interpretation as a transition of the effective bath chain.
At the non-Fermi-liquid critical point, the EE develops an unstable plateau.
The $\Lambda$ dependence of the critical-point EE is similar to that of the overscreened two-channel Kondo model, supporting universal behavior within the same non-Fermi-liquid universality class.
\end{abstract}
\maketitle

\paragraph{Introduction.}
Entanglement has become a powerful concept for characterizing quantum many-body phases. When the system is divided into two subsystems $A$ and $B$, they are entangled if the total state cannot be written as a simple tensor product of the states of these subsystems.
Notably, different kinds of bipartition lead to rather different entanglement; \eg a free-fermion state is a trivial tensor product state in momentum space but can possess super-area-law entanglement in real space \cite{wolf_violation_2006,gioev_entanglement_2006,swingle_entanglement_2010}. Most previous studies of entanglement focus on real-space entanglement. For example, the Li-Haldane conjecture \cite{li_entanglement_2008} relates the real-space entanglement spectrum to the edge excitations; for a one-dimensional system with conformal symmetry, which is common at the critical point, the real-space entanglement entropy is determined by the central charge \cite{calabrese_entanglement_2004,calabrese_entanglement_2009}, and the entanglement spectrum is the same as the spectrum of boundary conformal field theory \cite{kabat_comment_1994,cardy_entanglement_2016,cho_universal_2017}; for a gapped state, the topological EE, which is defined by the constant part of EE besides the area law contribution, detects the topological order of the state \cite{kitaev_topological_2006,levin_detecting_2006}. Building on these and many other results, studies using real-space EE have achieved remarkable success in characterizing quantum phases and phase transitions.

In contrast, while momentum-space entanglement has also been extensively studied, its applications are relatively rare. One reason is that interactions are usually local in real space, not momentum space; momentum-space cuts therefore lack an area law, making many-body calculations less efficient, especially at strong coupling \cite{xiang_density-matrix_1996,nishimoto_application_2002,ehlers_entanglement_2015}. More fundamentally, many properties of entanglement rely on locality of interaction, making momentum-space entanglement less universal. Refs.~\cite{lundgren_momentum-space_2014,dora_distilling_2017} study one-dimensional systems whose universal momentum-space entanglement spectra exhibit features similar to those of the physical spectra or can be described by conformal field theory (CFT), but their results are restricted to the bipartition between the left- and right-moving modes rather than a more general momentum-space cut.

In this Letter, we focus on a rather different scenario: the bipartite entanglement of a quantum impurity model in energy space.
Since the bath energy labels bath momentum states, this construction is also a momentum-space bipartition, although only the bath electrons are partitioned in energy/momentum space while the impurity remains local.
Specifically, we divide the system into two subsystems: one containing the high-energy bath electrons and the impurity, and the other containing the low-energy bath electrons.
The bath energy mesh is discretized logarithmically, in direct analogy with the poor man's scaling approach to the Kondo problem \cite{anderson_poor_1970}, where high-energy bath electrons are progressively integrated out and the coupling between the impurity and the remaining low-energy bath electrons is renormalized.
We find that the resulting entanglement entropy characterizes phases and phase transitions in quantum impurity models, including the competition between local-singlet and Kondo-singlet phases, which lies beyond the conventional Landau paradigm.

\begin{figure}
    \centering
    \includegraphics[width=0.8\linewidth]{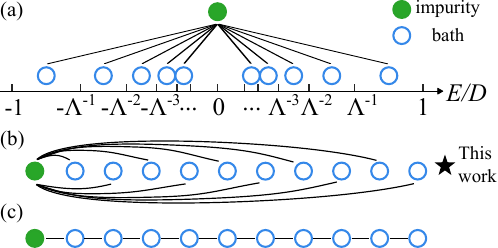}
    \caption{(a) Impurity-bath coupling and logarithmic bath discretization. (b,c) The star and chain representation, respectively. We use the star representation in this work. \label{fig:intro}}
\end{figure}

\paragraph{Model.} We consider an impurity coupled to a noninteracting bath,
\begin{align}
    \hH =  \hH_{f} +\hH_{c}  +  \hH_{cf}\, ,
\end{align}
where $c$ and $f$ denote bath and impurity degrees of freedom. Only local bath combinations couple to the impurity. We therefore replace the microscopic bath by one-dimensional energy modes chosen to reproduce those local bath correlators. For Anderson models this object is the hybridization function; for Kondo models it is the local bath field entering the spin density.

For the $N$-orbital Anderson impurity model (NoAIM), let $\alpha$ and $s$ denote the orbital and spin labels, respectively. We take $\hH_f
    =
    \sum_{\alpha s} \epsilon_{f,\alpha s}\hat n_{f,\alpha s}
    +\frac{U}{2}
    \sum_{(\alpha,s)\ne(\beta,s')}
    \hat n_{f,\alpha s}\hat n_{f,\beta s'} ,$
where $\epsilon_{f,\alpha s}$ is the on-site energy and $\hat n_{f,\alpha s}$ is the impurity occupation. The total Hamiltonian can always be written as
\begin{align}
    \hH_{\rm AIM}
    &=
    \hH_f
    +\sum_a\int_{-D}^{D}{\rm d}\varepsilon\,
    \varepsilon c^\dagger_{\varepsilon a}c_{\varepsilon a}
    \nonumber\\
    &\quad
    +\sum_{\alpha s,a}\int_{-D}^{D}{\rm d}\varepsilon\,
    \left[v_{\alpha s,a}(\varepsilon)f^\dagger_{\alpha s} c_{\varepsilon a}+h.c.\right],
\end{align}
where $D$ is the bath energy cutoff (half-bandwidth), $a$ labels the effective bath channels, and $[\Delta(\varepsilon)]_{\alpha s,\beta s'}=\sum_a v_{\alpha s,a}(\varepsilon)v^*_{\beta s',a}(\varepsilon)$ defines the hybridization function. This representation can capture all impurity physics for any kind of bath because any bath with the same $\Delta(\omega)$ gives the same Anderson impurity problem \cite{bulla_numerical_2008}.

For the $N$-channel Kondo model (NCK), the spin-$S_{\rm imp}$ impurity couples instead to the local bath spin density through $\hH_{cf}
    \!=\!J_K\hat{\mathbf S}_f\!\cdot\!\sum_{\alpha=1}^{N}\hat{\mathbf S}_{c,\alpha}(0)$, where $J_K\!>\!0$ , $
    \hat{\mathbf S}_{c,\alpha}(0)\!
    =\!\frac{1}{2}\sum_{ss'}\psi^\dagger_{\alpha s}(0){\bm\sigma}_{ss'}\psi_{\alpha s'}(0)$, where ${\bm\sigma}$ is the vector of Pauli matrices, $\alpha$ labels the channel, and $\psi_{\alpha s}(0)$ is the fermionic annihilation operator at the impurity site. $\hat{\mathbf S}_f$ is the spin operator of the impurity. The energy representation is obtained by rewriting the local field as $\psi_{\alpha s}(0)=\int_{-D}^{D}{\rm d}\varepsilon\,\sqrt{\nu(\varepsilon)}c_{\varepsilon\alpha s}$ where $\nu(\varepsilon)$ denotes the bath density of states. We use $S_{\rm imp}=1/2$ unless otherwise specified.

We then use a logarithmic mesh with discretization parameter $\Lambda>1$ to divide the bath into $L$ shells, where $i=1,\ldots,L$ labels the shell and $\eta=\pm$ its positive- or negative-energy branch (\cref{app:discretization}).
For the flat bath $\nu(\varepsilon)\!=\!1$, $\hH_c\!\to\!\sum_{i\eta a}\eta\varepsilon_i \hat n_{i\eta a}$, where $\hat n_{i\eta a}=c^\dagger_{i\eta a}c_{i\eta a}$, $\varepsilon_i=D(1+\Lambda^{-1})\Lambda^{-(i-1)}/2$, and $d_i=\Lambda^{-i+1}(1-\Lambda^{-1})$ is the dimensionless width of the $i$th logarithmic energy shell. For $[\Delta(\varepsilon)]_{\alpha s,\beta s'}=\Delta_0\delta_{\alpha\beta}\delta_{ss'}$, the Anderson coupling becomes $\int_{i,\eta}{\rm d}\varepsilon\,\sqrt{\Delta_0}\,c_{\varepsilon\alpha s}\to\sqrt{D\Delta_0d_i}\,c_{i\eta\alpha s}$; in the Kondo model, $\psi_{\alpha s}(0)\to\sum_{i\eta}\sqrt{Dd_i}\,c_{i\eta\alpha s}$.
The basis $c_{i\eta a}$ is known as the {\it star representation}.
A subsequent Lanczos procedure \cite{wilson_renormalization_1975,krishna-murthy_renormalization-group_1980-1,krishna-murthy_renormalization-group_1980,bulla_numerical_2008} maps the bath Hamiltonian to tridiagonal form, giving the (Wilson) {\it chain representation} with nearest-neighbor hoppings, the standard starting point of the numerical renormalization group (NRG) \cite{wilson_renormalization_1975,krishna-murthy_renormalization-group_1980-1,krishna-murthy_renormalization-group_1980,bulla_numerical_2008}.

In both representations, bath sites are organized by energy scale, but the structures of their entanglement are different. The bath sites in the star representation are eigenmodes of the decoupled bath and become entangled only through their coupling to the impurity. However, in chain representation, the bath sites are entangled even without the impurity. One may try to remove this bath contribution by subtracting the entropy of the impurity-free chain. However, below the Kondo scale this impurity contribution vanishes \cite{wagner_long-range_2018}, so its fixed-point value cannot distinguish Fermi-liquid phases with and without Kondo screening. We therefore prefer the star representation, which we call ``energy space'', and, as shown below, whose fixed-point EE takes distinct universal values in these phases.
We order the impurity first, followed by bath sites with energies $-\varepsilon_1,\varepsilon_1,\ldots,-\varepsilon_L,\varepsilon_L$ (\cref{fig:intro}). A cut to the left of $-\varepsilon_i$ separates the impurity and higher-energy modes from the lower-energy bath, and the corresponding EE is denoted by $S_\Lambda(\varepsilon_i)$. 

We solve the model with density matrix renormalization group (DMRG), where the sites are ordered as descibed above.
DMRG is implemented using the ITensor Julia libraries \cite{ITensor,ITensor-r0.3,corbett_scaling_2025}; the Hamiltonian admits a compact MPO representation, whose bond dimension is independent of system size, although the Hamiltonian is nonlocal in energy space (\cref{app:MPO}).

\begin{figure}
    \centering
    \includegraphics[width=\linewidth]{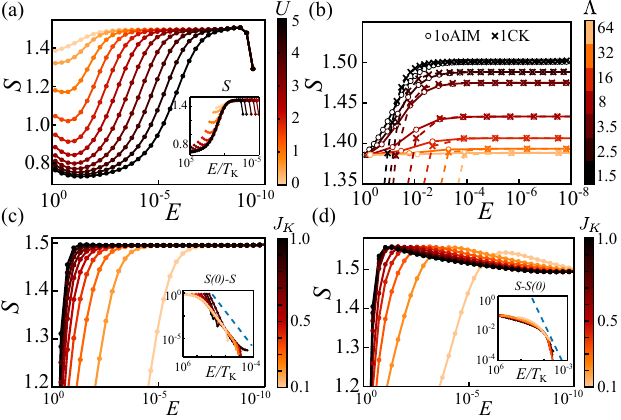}
    \caption{(a) $S_\Lambda(E)$ for the Anderson model with $U=0,0.5,\ldots,5$ (bright to dark), $\Delta_0=0.05$, $D=1$, and $\Lambda=2$. Inset: the same data versus $E/T_K$. (b) $S_\Lambda(E)$ for the Anderson and Kondo models at different $\Lambda$, with $U=0$, $\Delta_0=0.05$, and $J_K=0.5$. (c,d) $S_\Lambda(E)$ for the fully screened and underscreened Kondo models with $J_K=0.1,0.2,\ldots,1.0$ (bright to dark). Insets: $|S_\Lambda(0)-S_\Lambda(E)|$ versus $E/T_K$; the dashed blue lines indicate linear scaling with $E$. All DMRG calculations use maximum bond dimension $\chi=1600$. \label{fig:FL}}
\end{figure}
\paragraph{The Kondo phase.}
We demonstrate the universality of energy-space EE using the Anderson impurity model and the fully screened and underscreened Kondo models, all with Fermi-liquid fixed points.

We first consider the $N=1$ Anderson impurity model with particle-hole symmetry ($\varepsilon_f=-\frac{1}{2}U$) and a constant hybridization function $\Delta(\omega)=\Delta_0$. The energy dependence of $S_\Lambda(E)$ resolves the physics at different energy scales, as shown in \cref{fig:FL}(a).

At the highest cut, $S_\Lambda$ measures the entanglement between the bare impurity and the bath. It decreases from $2\ln 2$ to $\ln 2$ as $U$ increases: in the $U=0$ limit, the impurity reduced density matrix is maximally mixed over its four-dimensional Hilbert space, whereas in the large-$U$ limit, charge fluctuations are suppressed and it is maximally mixed within the spin-$\frac{1}{2}$ doublet. This reveals the suppression of charge fluctuations on the impurity.

As $E$ goes to zero, $S_\Lambda(E)$ saturates to a value independent of $U$ and slightly larger than $2\ln 2$, as shown in \cref{fig:FL}(a). The deviations of $S_\Lambda(E)$ in the lowest few energies are due to the finite-size effects and should be ignored, and we will hereafter denote the converged plateau value by $S_\Lambda(0)$. We can identify the Kondo temperature $\tk$ from the energy scale at which $S_\Lambda(E)$ approaches $S_\Lambda(0)$ and $\tk$ decreases exponentially with $U$ as expected \cite{Hewson_1993}. $S_\Lambda$ as a function of $E/\tk$ collapses into a single line for different $U$, verifying the well-known result that $\tk$ is the only energy scale in the low-energy limit.

\begin{figure}
    \centering
    \includegraphics[width=\linewidth]{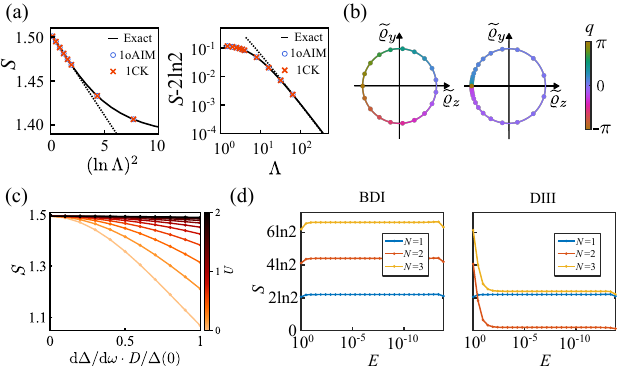}
    \caption{(a) Comparison of $S_\Lambda(0)$ at different $\Lambda$ for the Anderson impurity model, the Kondo model obtained from DMRG, and the exact value from directly diagonalizing the free-fermion Hamiltonian. The dashed lines show the asymptotic behaviors at $\Lambda\to 1$ and $\Lambda\to\infty$ limits. (b) $(\td{\rho}_{\Lambda,z}(q),\td{\rho}_{\Lambda,y}(q))$ when $\Lambda\to\infty$ (left panel) and $\Lambda\to1$ (right panel). The color represents $q$ and the points are sampled on uniform $q$ meshes. We use maximum bond dimension $\chi=1600$ for DMRG calculations in (a). (c) The fixed-point entropy $S$ under bath particle-hole asymmetry for the one-orbital Anderson model with $\Delta(\omega)=\Delta(0)[1+x\omega/D]$, plotted versus the dimensionless slope $x$. $\Delta(0)=0.05$, $D=1$, and $\Lambda=2$. The color denotes $U=0,0.2,\ldots,2.0$. (d) $S_\Lambda(E)$ for representative quadratic impurity models in classes BDI ($\mathcal{T}^2=1,\mathcal{C}^2=1$) and DIII ($\mathcal{T}^2=-1,\mathcal{C}^2=1$). Here $N$ labels the number of spinful orbitals. The corresponding AZ symmetry constraints and representative constructions are given in \cref{app:sym-topo}. \label{fig:topo}}
\end{figure}

We next consider the fully screened 1CK model ($N=2S_{\rm imp}=1$), whose $S_\Lambda(E)$ approaches the same fixed-point value as the Anderson model, as shown in \cref{fig:FL}(b). Comparing the full $\Lambda$ dependence in \cref{fig:FL}(b) and \cref{fig:topo}(a) further shows that this agreement persists for every $\Lambda$, confirming universality across these Fermi-liquid models.

$S_\Lambda(E)$ in the underscreened Kondo model (1CK with $S_{\rm imp}\!=\!1$) reaches the same fixed-point value, but exhibits a much longer finite-energy tail.
As shown in \cref{fig:FL}(c)(d), $|S_\Lambda(E)\!-\!S_\Lambda(0)| \!\propto\! E$ for the fully screened Kondo model, whereas the decay is substantially slower in the underscreened case.
This is consistent with its singular Fermi-liquid nature \cite{coleman_singular_2003,koller_singular_2005}: the low-energy fixed point is asymptotically Fermi-liquid-like, but the residual unscreened moment induces logarithmic corrections at finite energies.
Therefore, both the fixed-point value and the low-energy approach of $S_\Lambda(E)$ can be used to characterize the nature of the impurity fixed point.

\paragraph{Emergent symmetries and topological classification.} In the Kondo phase, the impurity is renormalized into a Fermi liquid in the low-energy limit. The fixed point can therefore be represented by the $U\!=\!0$ Anderson impurity model, consistent with the independence of $S_\Lambda(0)$ from $U$ and $J_K$. We find that emergent symmetries at this Fermi-liquid fixed point gives a topological classification of the fixed-point wavefunction in energy space, explaining the universal value of $S_\Lambda(0)$.

First, the fixed point is scale invariant, as reflected in the low-energy plateau of $S_\Lambda(E)$ in \cref{fig:FL}. Because the bath shells are logarithmically spaced, scale transformations act as translations along the energy-space chain. For the Anderson impurity model, the single-particle density matrix $\rho_{i\eta,j\eta'} = \langle c^\dagger_{i\eta} c_{j\eta'}\rangle$ satisfies $\rho_{i,\eta;i',\eta'}=\rho_{i+r,\eta;i'+r,\eta'}$ away from the UV and IR cutoffs (\cref{app:Anderson}), provided the hybridization function is smooth near the Fermi level. This allows us to treat the fixed-point wavefunction as a one-dimensional translation-invariant free-fermion state.

The second ingredient is chiral symmetry of the low-energy fixed-point Hamiltonian. At the microscopic particle-hole-symmetric point, the renormalized impurity level lies exactly at the Fermi energy, $\varepsilon_f^*=0$. Away from this point, $\lvert\varepsilon_f^*\rvert$ generally remains of order $T_K$ provided the impurity valence is closed to 1, so the level can still lie close to zero when $T_K$ is small. The fixed-point quasiparticles probe only the window $\lvert\omega\rvert\lesssim T_K$. If the hybridization function is smooth over this window, then $\Delta(\omega)\simeq\Delta(-\omega)\simeq\Delta(0)$. Together, $\varepsilon_f^*\simeq0$ and an approximately even hybridization give the fixed-point Hamiltonian an approximate chiral symmetry that exchanges the positive- and negative-energy bath branches. We use $\tau_\mu$ for the Pauli matrices acting in the energy branch basis $(\eta=+,\eta=-)$. In this basis, the chiral operator is $\tau_x$ and anticommutes with the effective single-particle Hamiltonian. At finite impurity-level or bath asymmetry, the fixed-point entropy approaches its symmetric value as $U$ increases and $T_K$ decreases, as shown in \cref{fig:onsite-asym-appendix} and \cref{fig:topo}(c), respectively.

For the Anderson models considered here, combining physical time reversal with an $SU(2)$ spin rotation gives an effective symmetry $\mathcal T'\!=\!\sigma_0K$, where $K$ denotes complex conjugation and $(\mathcal T')^2\!=\!1$. Whenever the low-energy chiral symmetry described above is present, its combination with $\mathcal T'$ places the fixed point in class BDI of the Altland-Zirnbauer classification \cite{zirnbauer_riemannian_1996,altland_nonstandard_1997,schnyder_classification_2008,ryu_topological_2010,chiu_classification_2016,ludwig_topological_2016}, with a $\mathbb Z$ invariant.

In detail, we define the scale Fourier mode $[\td{\rho}_\Lambda(q)]_{\eta\eta'} = \sum_{r}\rho_{i\eta,(i+r)\eta'} e^{\ii qr}$ with $q\in[-\pi,\pi)$. The symmetries constrain  $
    \td{\rho}_\Lambda(q)= \frac{1}{2}\tau_0 + \td{\rho}_{\Lambda,y}(q)\tau_y +  \td{\rho}_{\Lambda,z}(q)\tau_z .
$
Because the fixed-point ground state is a pure free-fermion state, the eigenvalues of $\td{\rho}_\Lambda(q)$ are $0$ or $1$, and hence $|\td{\rho}_{\Lambda,z}(q)|^2+|\td{\rho}_{\Lambda,y}(q)|^2=1/4$. The winding of $q\mapsto(\td{\rho}_{\Lambda,z}(q),\td{\rho}_{\Lambda,y}(q))$ as $q$ changes from $-\pi$ to $\pi$ is the BDI topological invariant. We find winding number one for all $\Lambda$. In the $\Lambda\to \infty$ limit, the analytic calculation in \cref{app:analytic} gives
\begin{align}
    \td{\rho}_\infty(q)=\frac{1}{2}\tau_0+\frac{1}{2}(\tau_z\cos q+\tau_y\sin q),
\end{align}
which is the same as the Hamilonian of Su-Schrieffer-Heeger (SSH) chain \cite{su_solitons_1979,shen_topological_2017} with only nearest neighbouring inter-unit cell hopping. Intuitively, $\Lambda\to\infty$ is the strongest coarse graining of energy space, so the fixed point has the shortest correlation length while remaining topologically nontrivial. The semi-infinite chain then has one boundary mode and gives $S_\infty(0)=\ln2$ per flavor. For finite $\Lambda$, longer-range matrix elements increase the effective correlation length and produce universal positive corrections to the boundary entropy.

In the $\Lambda\to1$ limit, the logarithmic energy spacing $\delta_\Lambda=\ln\Lambda$ between neighboring bath sites vanishes, and the energy-space chain approaches a continuum model. In the continuum coordinate $i\delta_\Lambda\simeq\ln(D/\varepsilon_i)$, correlations extend over a fixed logarithmic-energy window rather than a fixed number of bath sites. As derived in \cref{app:analytic}, we obtain
{\small
\begin{align}
    \td{\rho}_\Lambda(q)\simeq \frac{1}{2}\tau_0 + \frac{1-\sinh^2(\pi q/\ln\Lambda)}{2\cosh^2(\pi q/\ln\Lambda)}\tau_z + \frac{\sinh(\pi q/\ln\Lambda)}{\cosh^2(\pi q/\ln\Lambda)}\tau_y\, .
\end{align}
}
The full winding concentrates within the region $q=O(\ln\Lambda)$, as illustrated in the right panel of \cref{fig:topo}(b). The EE is obtained from the eigenvalues of $\rho$ restricted to the semi-infinite chain. The topological boundary mode contributes $\ln2$ per flavor, while the positive correction comes from continuum correlations across the logarithmic-energy window.

The same argument extends to multiorbital and other symmetry classes. For the $N$-orbital Anderson model in the BDI setting above, the fixed-point wavefunction
is composed of $2N$ copies of the SSH chain, giving a topological invariant $2N$ and an entropy $S_\Lambda(0)=2N\ln2$ plus $\Lambda$-dependent corrections.
More general impurity models can be obtained by imposing the corresponding symmetry constraints on the hybridization function $\Delta(\omega)$ and on the local single-particle impurity Hamiltonian $H_f$, so that all AZ classes can be realized and $S_\Lambda(0)$ reflects the one-dimensional topological classification of the fixed-point Hamiltonian (\cref{app:sym-topo}). Representative BDI and DIII results are shown in \cref{fig:topo}(d), with classifications $\mathbb{Z}$ and $\mathbb{Z}_2$, respectively. The DIII representative describes Majorana Kramers pairs coupled to time-reversal-related Majorana bath modes, with particle-hole symmetry interpreted as the BdG/Majorana constraint rather than ordinary number-conserving charge conjugation, as discussed in \cref{app:sym-topo}.

\paragraph{Quantum phase transition and non-Fermi-liquid behavior.} We further demonstrate that energy-space entanglement can also characterize quantum phase transitions. In particular, it captures a transition between two Fermi-liquid phases without symmetry breaking, beyond the traditional Landau framework.

We consider the two-orbital Anderson model and add an inter-orbital antiferromagnetic coupling $J_S\geq 0$. The Hamiltonian is $\hH_{\rm AIM}+\hH_J$ where $\hH_{\rm AIM}$ is the the same as before and we also consider particle-hole symmetry case $\varepsilon_f=-\frac{3}{2}U$ and a constant hybridization function, and $\hH_J = J_S \hat{\mathbf S}_{f,1}\cdot \hat{\mathbf S}_{f,2}$, $\hat{\mathbf S}_{f,\alpha}=\frac{1}{2}\sum_{ss'}f^\dagger_{\alpha s}{\bm \sigma}_{ss'}f_{\alpha s'}$. 
\begin{figure}
    \centering
    \includegraphics[width=\linewidth]{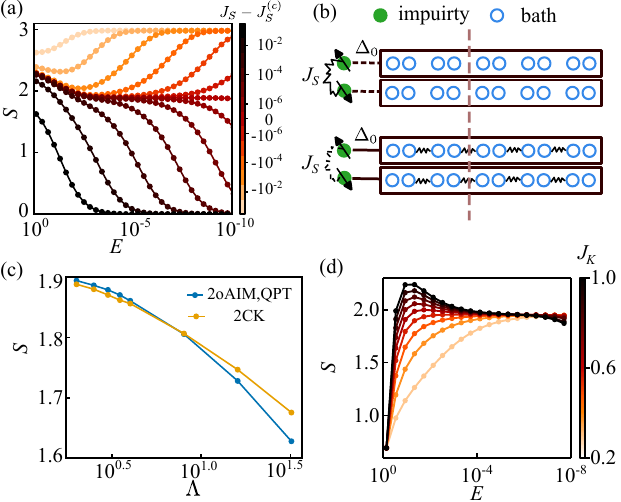}
    \caption{(a) $S_\Lambda(E)$ for the 2oAIM with $J_S$ close to $J_S^{(c)}$. (b) Bath-topology schematic for the local-singlet--Kondo-singlet transition. The top and bottom panels denote the local-singlet and Kondo phases, while dashed and solid black bonds denote weak and strong couplings. The two boxes denote orbitals, and each unit cell contains the two energy-branch bath modes at a given scale, up to a local rotation between positive and negative energies. The vertical dashed line marks the entanglement cut. (c) $S_\Lambda(0)$ versus $\Lambda$ at the QPT of 2oAIM and 2CK. (d) $S_\Lambda(E)$ for 3CK. Parameters are $\Lambda\!=\!2,U\!=\!0.5$, $\Delta_0\!=\!0.05,D\!=\!1$, and $J_S^{(c)}
\!=\!0.2349895$ in (a), and $\Lambda\!=\!2.5$ and $J_K\!=\!0.2,0.3,\ldots,1.0$ (bright to dark) in (d). The maximum DMRG bond dimensions are $\chi=1600$ in (a,c) and $\chi=12800$ in (d).\label{fig:NFL}}
\end{figure}

This model shares physics with the two-impurity Anderson model and their Kondo counterparts, which are extensively studied in previous works \cite{jayaprakash_two-impurity_1981,affleck_exact_1992,affleck_conformal-field-theory_1995,jones_study_1987,gan_solution_1995,leo_spectral_2004,fabrizio_nontrivial_2003,zarand_quantum_2006,mross_two-impurity_2008,mitchell_two-channel_2012,nishikawa_convergence_2012}.
At small $J_S$, the model is adiabatically connected to the $SU(4)$ Anderson model and flows to a Kondo fixed point, where the impurity and bath form a Kondo singlet. At large $J_S$, the impurity instead forms a local singlet and effectively decouples from the bath. Accordingly, \cref{fig:NFL}(a) shows that $S_\Lambda(0)$ remains at its $J_S=0$ value in the Kondo phase, twice the single-orbital value, and vanishes in the local-singlet phase. The star representation therefore distinguishes these two Fermi-liquid fixed points, whose fixed-point EEs coincide in the chain representation.

The emergent chiral symmetry discussed above provides a topological interpretation of this EE distinction. When this symmetry holds at low energies, the effective energy-space bath chain lies in the topological SSH phase at the Kondo fixed point, whereas at the local-singlet fixed point it reduces to isolated, unentangled bath modes in the trivial phase. The transition can then be viewed as a bath topological transition, as sketched in \cref{fig:NFL}(b). As discussed for $J_S=0$, weak particle-hole asymmetry can leave the emergent chiral symmetry intact, whereas sufficiently strong asymmetry breaks it and removes the topological distinction between the two phases.
Energy-space EE nevertheless always remain a useful diagnostic: the screened Kondo fixed point has nonzero entropy, while the decoupled local-singlet fixed point has zero entropy.

Near $J_S\!=\!J_S^{(c)}$, $S_\Lambda(E)$ develops a new plateau. Away from criticality, the low-energy flow eventually leaves the plateau, indicating that it corresponds to an unstable critical fixed point.
The critical point is a known non-Fermi liquid in the same universality class as the two-channel Kondo model. Consistently, their $S_\Lambda(0)$ curves show similar $\Lambda$ dependence in \cref{fig:NFL}(c), supporting the universality of the energy-space EE within this non-Fermi-liquid class. Extrapolation of the critical-point and two-channel Kondo data toward $\Lambda\to1$ suggests limiting values above $2\ln2+\ln\sqrt2$. The limiting EE at these two fixed points may contain a $2\ln2$ Fermi-liquid contribution together with an additional $\ln\sqrt2$ non-Fermi-liquid contribution \cite{gan_solution_1995,fabrizio_nontrivial_2003}, but this interpretation remains speculative. 

More generally, \cref{fig:NFL}(d) shows $S_\Lambda(E)$ for the overscreened 3CK model, whose plateau value is independent of $J_K$, further suggesting that the fixed-point energy-space EE is universal in NFL, with its value governed by the universality class rather than microscopic couplings.

\paragraph{Discussion.} In this work, we introduced an energy-space bipartition for logarithmically discretized quantum impurity models and computed the corresponding EE. The resulting entanglement profile resolves the RG flow in energy: Fermi-liquid fixed points give universal low-energy plateaus, while the local-singlet and Kondo-singlet phases are distinguished by different fixed-point EEs and bath topologies. The logarithmic discretization concentrates numerical resolution near the Fermi level, allowing exponentially small Kondo and crossover scales to be reached with a moderate number of bath shells, and turning fixed-point scale invariance into translation invariance along the energy-space chain.

Topological interpretations of the local-singlet--Kondo-singlet transition have also been discussed previously \cite{seki_topological_2017,curtin_fermi_2018,blesio_topological_2018,nishikawa_magnetic_2018}, but from a different perspective. There, the two Fermi liquids are distinguished by whether the relevant sum rule is obeyed and by the associated Luttinger integral, a Green's-function winding number \cite{seki_topological_2017}. In impurity problems, this invariant is defined from the impurity Green's function rather than the bath Green's function, so its relation to our energy-space bath topology remains unclear.

A natural next step is to extend energy-space EE to lattice models. In one dimension or isotropic model in higher dimension, this may be done by directly cutting the single-particle energy or momentum modes, in a construction related to field-theory studies of momentum-space entanglement \cite{balasubramanian_momentum-space_2012,flynn_momentum_2023,kong_momentum_2025}, although those works mainly rely on perturbative calculations. In higher dimensions, the Fermi-surface geometry makes such a direct construction less straightforward. An alternative route is through dynamical mean-field theory, where a lattice problem is represented by a self-consistent impurity model; the energy-space EE of this impurity problem may then serve as a diagnostic of the lattice phase. This perspective is also connected to the auxiliary-chain description of the Hubbard-model Mott transition as a topological transition of an effective SSH model \cite{sen_mott_2020}.

\begin{acknowledgments}
We thank Y.-J. W. and S.-S. L. for helpful discussions. Z.-D.~S. and G.-D.~Z.~were supported by National Natural Science Foundation of China (Nos.~12274005 and 12521006), National Key Research and Development Program of China (No.~2021YFA1401900), and Quantum Science and Technology-National Science and Technology Major Project (No.~2021ZD0302403).
\end{acknowledgments}

\bibliography{ref}
\clearpage
\onecolumngrid
\appendix
\tableofcontents
\section{Details of the bath representation \label{app:discretization}}
In this section, we summarize the standard procedure for rewriting the bath in one-dimensional energy space and logarithmically discretizing it in NRG \cite{wilson_renormalization_1975,krishna-murthy_renormalization-group_1980,krishna-murthy_renormalization-group_1980-1,bulla_numerical_2008}, making this work self-contained.
\subsection{Anderson model}
The original Hamiltonian of the Anderson impurity model is
\begin{align}
    \hH = \hH_{f}+\hH_{cf}+\hH_c
\end{align}
where $\hH_{cf} = N_s^{-1/2}\sum_{\kk\alpha s}V_\kk c^\dagger_{\kk\alpha s}f_{\alpha s} +h.c.,\,\hH_c = \sum_{\kk\alpha s}\varepsilon_{\kk} c^\dagger_{\kk\alpha s}c_{\kk\alpha s}$, and $N_s$ is the number of discrete bath momentum states. Here $\alpha$ denotes the orbital index and $s$ denotes the spin index. We focus on coupling that is diagonal in $\alpha,s$ here and leave the generic discussion to \cref{app:sym-topo}. The effect of the bath on the impurity is solely determined by the hybridization function defined by
\begin{align}
    \Delta(\omega) = \frac{1}{N_s}\sum_\kk |V_\kk|^2 \delta(\omega-\varepsilon_\kk)\,.
\end{align}
For the flat hybridization used in the main text, we take $\Delta(\omega)=\Delta_0$ for $|\omega|<D$, where $\Delta_0$ denotes the height of the spectral hybridization function. In the convention $\Gamma(\omega)=\pi\Delta(\omega)$, the corresponding level broadening is $\Gamma_0=\pi\Delta_0$.

The matrix factorization introduced in the main text reduces to a scalar square root in this flavor-diagonal case. For the diagonal hybridization function considered in this subsection, one may choose the following Hamiltonian to reproduce it:
\begin{align}
    \hH = \hH_{f}+ \sum_{\alpha s}\int_{-D}^{D}{\rm d}\varepsilon\,\varepsilon  c^\dagger_{\varepsilon \alpha s} c_{\varepsilon \alpha s} + \sum_{\alpha s}\int_{-D}^{D}{\rm d}\varepsilon\, \sqrt{\Delta(\varepsilon)} (c^\dagger_{\varepsilon \alpha s}f_{\alpha s}+h.c.)
\end{align}
After logarithmic discretization, the $i$th positive and negative energy intervals are $\int_{i,+} = \int_{D\Lambda^{-i}}^{D\Lambda^{-i+1}}$ and $\int_{i,-} = \int_{-D\Lambda^{-i+1}}^{-D\Lambda^{-i}}$. We define the dimensionless interval width
\[
d_i=\Lambda^{-i+1}(1-\Lambda^{-1})
\]
and the shell fermion $c_{i,\pm} = \frac{1}{\sqrt{Dd_i}}\int_{i,\pm} {\rm d}\varepsilon c_{\varepsilon}$. We refer to the pair of positive- and negative-energy intervals with the same $i$ as the $i$th logarithmic energy shell. In this case, $c_{\varepsilon} \approx \frac{1}{\sqrt{Dd_i}}c_{i,\pm}$ when $\varepsilon \in \pm (D\Lambda^{-i+1},D\Lambda^{-i})$, respectively. The final Hamiltonian becomes
\begin{align}
    \hH = \hH_f + \sum_{\alpha s}\sum_{\eta=\pm}\sum_{i=1}^L t_{i,\eta}( c^\dagger_{i\eta \alpha s} f_{\alpha s} +h.c.) +  \sum_{\alpha s}\sum_{\eta=\pm}\sum_{i=1}^L \eta \varepsilon_i  c^\dagger_{i\eta \alpha s} c_{i\eta \alpha s}\label{eq:HAIM-dis}
\end{align}
where $\varepsilon_i = D(1+\Lambda^{-1})\Lambda^{-(i-1)}/2$ and $t_{i,\pm}^2 = \int_{i,\pm} {\rm d}\varepsilon \Delta(\varepsilon)$.
\subsection{Kondo model}
The Hamiltonian for a general Kondo model is
\begin{align}
    \hH = \sum_{\kk \alpha s} \varepsilon_\kk c^\dagger_{\kk\alpha s} c_{\kk\alpha s}  + J_K{\bm S}_c(0)\cdot {\bm S}_f
\end{align}
where the local bath spin density is
\begin{align}
    {\bm S}_c(0)
    =\frac{1}{2}\sum_\alpha\sum_{ss'}
    \psi^\dagger_{\alpha s}(0){\bm \sigma}_{ss'}\psi_{\alpha s'}(0),
\end{align}
and the local bath field is represented in energy space as
\begin{align}
    \psi_{\alpha s}(0)
    =
    \int_{-D}^{D}{\rm d}\varepsilon\,
    \sqrt{\nu(\varepsilon)}c_{\varepsilon\alpha s},
\end{align}
with $\{c_{\varepsilon\alpha s},c^\dagger_{\varepsilon'\beta s'}\}=\delta(\varepsilon-\varepsilon')\delta_{\alpha\beta}\delta_{ss'}$. Here $\nu(\varepsilon)$ denotes the bath density of states per channel and spin. Throughout this work, we use $\nu(\varepsilon)=1$ for $|\varepsilon|<D$. The factor $\sqrt{\nu(\varepsilon)}$ comes from the density of momentum states in the local field. This differs from the Anderson energy representation, where the coupling contains $\sqrt{\Delta(\varepsilon)}$ because the microscopic hybridization amplitudes are also included.
The Kondo Hamiltonian in the energy representation is therefore
 \begin{align}
     \hH = \sum_{\alpha s}\int^D_{-D} {\rm d}\varepsilon \varepsilon c^\dagger_{\varepsilon \alpha s}c_{\varepsilon \alpha s} + \frac{1}{2}J_K\sum_{\alpha ss'}\int_{-D}^D{\rm d}\varepsilon \int_{-D}^D{\rm d}\varepsilon' \sqrt{\nu(\varepsilon)\nu(\varepsilon')}(c^\dagger_{\varepsilon\alpha s}{\bm \sigma}_{ss'}c_{\varepsilon'\alpha s'})\cdot {\bm S}_f
 \end{align}
 Discretizing the bath, we have
 \begin{align}
    \hH =  \sum_{\alpha ss'} J_K{\bm S}_f \cdot \frac{1}{2} {\bm \sigma}_{ss'}
    \sum_{i,i'=1}^{L}\sum_{\eta,\eta'=\pm}
    t_{i\eta} t_{i'\eta'}c^\dagger_{i\eta\alpha s}c_{i'\eta'\alpha s'}
    +\sum_{\alpha s}\sum_{\eta=\pm}\sum_{i=1}^L
    \eta \varepsilon_i c^\dagger_{i\eta \alpha s} c_{i\eta \alpha s}
    \label{eq:HK-dis}
\end{align}
where $t_{i,\pm}^2 = \int_{i,\pm} {\rm d}\varepsilon \nu(\varepsilon)$. For the $S_{\rm imp}=\frac{1}{2}$ case, the Kondo Hamiltonian can also be obtained by making a Schrieffer-Wolff transformation from the Anderson Hamiltonian in the large-$U$ limit, and the factorized bath weights in \cref{eq:HK-dis} follow from the discretized local field above.

\section{Analytical results \label{app:analytic}}
\subsection{General formulation for free fermions}
For free fermions, the reduced density matrix can be obtained from the correlation function \cite{peschel_calculation_2003}. Considering a free-fermion system whose many-body reduced density matrix in a certain subsystem is $\hat\rho$, the two-point correlation function, or single-particle density matrix in this subsystem, is $\rho_{ij} = \tr(\hat\rho c^\dagger_{i}c_j)$.
By Wick's theorem, the reduced density matrix of the subsystem is Gaussian, so it is fully determined by the eigenvalues $n_r$ of the restricted single-particle density matrix $\rho_{ij}$. These $n_r$ can be viewed as the occupation probabilities of independent entanglement modes. Applying the von Neumann entropy to these modes gives
\begin{align}
    S = \sum_r\left( -n_r\ln n_r- (1-n_r)\ln(1-n_r)\right)
\end{align}

\subsection{Application to Anderson impurity model \label{app:Anderson}}
Here we consider the Anderson impurity model with a constant hybridization function at $U=0$, for which the problem is exactly solvable.
For the logarithmic discretization scheme, the Hamiltonian is
\begin{align}
    H = \begin{pmatrix}
        0 & t & t & t\Lambda^{-\frac{1}{2}} & t\Lambda^{-\frac{1}{2}} & t\Lambda^{-1} & t\Lambda^{-1} & \cdots &t\Lambda^{-\frac{L-1}{2}} & t\Lambda^{-\frac{L-1}{2}} \\
        t & \varepsilon &  &&& \cdots &&&&0\\
        t &  &-\varepsilon &&& \cdots &&&&0 \\
        t\Lambda^{-\frac{1}{2}}  &  &  &\varepsilon \Lambda^{-1}  && \cdots &&&&0 \\
        t\Lambda^{-\frac{1}{2}}  &  &  &  &-\varepsilon \Lambda^{-1}& \cdots &&&&0
    \end{pmatrix}\, ,
\end{align}
where $\varepsilon=\varepsilon_1,t=t_{1,+}=t_{1,-}$. The single-particle Hamiltonian $H$ respects a chiral symmetry represented by $U_S$,
\begin{align}
    U_SHU_S^{-1} = -H,\qquad U_S = \begin{pmatrix}
        -1 & 0 & 0 &  \cdots \\
        0 & \tau_x & 0 &  \cdots \\
        0 & 0 & \tau_x &  \cdots \\
        \vdots&\vdots&\vdots&\ddots
    \end{pmatrix}\, .
\end{align}
Physically, this chiral symmetry comes from the combination of the particle-hole symmetry and time-reversal symmetry of the model.
Because the Hamiltonian has chiral symmetry and odd dimension, it necessarily has an $E_0=0$ eigenstate. The other eigenstates come in pairs, and we label their positive absolute energies in descending order by $E_1,E_2,\cdots,E_L$.

To distinguish the original bath basis from the diagonal basis, we use $(i,\eta)$ and $(j,\eta')$ for bath states, where $i,j$ are logarithmic-shell indices and $\eta,\eta'=\pm$ label the positive- and negative-energy branches. A single-particle eigenstate is instead labelled by $(m,\zeta)$, where $m$ enumerates the absolute energies and $\zeta=\pm$ labels the sign of its eigenenergy, $E_{m\zeta}=\zeta E_m$. The unitary matrix that diagonalizes $H$ is written as
\begin{align}
    u_{i\eta,m\zeta}=\langle i,\eta|m,\zeta\rangle ,
\end{align}
and its impurity component is denoted by $u_{f,m\zeta}$. Thus the eigenvector with eigenvalue $\zeta E_m$ is
\begin{align}
    \mathbf{u}_{m\zeta}
    =
    \left(
    u_{f,m\zeta},
    u_{1+,m\zeta},u_{1-,m\zeta},
    \ldots,
    u_{L+,m\zeta},u_{L-,m\zeta}
    \right)^T .
\end{align}
Writing the single-particle Hamiltonian in impurity--bath block form,
\begin{align}
    H=
    \begin{pmatrix}
        0 & V^\dagger\\
        V & H_c
    \end{pmatrix},
\end{align}
and separating an eigenvector into its impurity and bath components as $\mathbf{u}_{m\zeta}=(\psi_f,\psi_c)^T$, the eigenvalue equation becomes
\begin{align}
    V^\dagger \psi_c = E\psi_f, V\psi_f + H_c\psi_c = E\psi_c\, .
\end{align}
Therefore, we have
\begin{align}
    E = V^\dagger(E-H_c)^{-1}V
\end{align}
and then
\begin{align}
    E = \sum_{i=1}^{L} \left(
    \frac{t^2}{\Lambda^{i-1}E-\varepsilon}
    +\frac{t^2}{\Lambda^{i-1}E+\varepsilon}
    \right)
\end{align}
We know that $E_m$ is of order $\varepsilon\Lambda^{-m+1/2}$. For shells $i\gg m$, $\Lambda^{i-1}E_m$ is large and the corresponding terms are small; for $i\ll m$, $\Lambda^{i-1}E_m$ is small and the two terms approximately cancel. For a bulk eigenstate, we write $E_m=\Lambda^{-(m-1)}z$ and set $r=i-m$. Taking $m\to\infty$ and $L-m\to\infty$ at fixed $\Lambda>1$, the summation range can be extended to $r=-\infty,\ldots,\infty$, while the left-hand side vanishes.
The equation is then
\begin{align}
    0 = \sum_{r=-\infty}^{\infty}
    \left(\frac{t^2}{\Lambda^r z-\varepsilon}
    +\frac{t^2}{\Lambda^r z+\varepsilon}\right)
    =2t^2z\sum_{r=-\infty}^{\infty}
    \frac{\Lambda^r}{\Lambda^{2r}z^2-\varepsilon^2}.
\end{align}
For any fixed $\Lambda>1$, this equation has the exact bulk solution $z=\Lambda^{-\frac{1}{2}}\varepsilon$, because the $r$th term and the $(1-r)$th term cancel pairwise:
\begin{align}
    \frac{\Lambda^{r}}{(\Lambda^{2r-1}-1)\varepsilon^2}
    +\frac{\Lambda^{1-r}}{(\Lambda^{1-2r}-1)\varepsilon^2}=0.
\end{align}

As $\psi_c = (E-H_c)^{-1} V\psi_f$ and
the corresponding eigenenergy is $\zeta E_m=\zeta\varepsilon\Lambda^{-m+\frac{1}{2}}$, we have
\begin{align}
    u_{i\eta,m\zeta}
    =
    \frac{t\Lambda^{-\frac{i-1}{2}}u_{f,m\zeta}}
    {(\zeta\Lambda^{-m+\frac{1}{2}}-\eta\Lambda^{-(i-1)})\varepsilon}\, .
\end{align}
Normalizing it, we obtain
\begin{align}
     u_{i\eta,m\zeta}
     =
     \frac{\alpha\Lambda^{\frac{i-m}{2}}}
     {\zeta\Lambda^{i-m-\frac{1}{2}}-\eta}
\end{align}
where the constant $\alpha$ is defined by
$u_{f,m\zeta}=(\alpha\varepsilon/t)\Lambda^{-(m-1)/2}$ and satisfies
\begin{align}
    \alpha^2 \sum_{r=-\infty}^\infty
    \left[
    \frac{\Lambda^r}{(\Lambda^{r-\frac{1}{2}}-1)^2}
    +\frac{\Lambda^r}{(\Lambda^{r-\frac{1}{2}}+1)^2}
    \right]
    +\alpha^2\frac{\varepsilon^2}{t^2}\Lambda^{-(m-1)}=1.
\end{align}
We focus on $m\gg1$, for which the second term is negligible.
$u_{i\eta,m\zeta}$ has maximal absolute value at the bath components $(i,\eta)=(m,\zeta)$ and $(m+1,\zeta)$:
\begin{align}
    |u_{m\zeta,m\zeta}|
    =
    |u_{(m+1)\zeta,m\zeta}|
    =
    \frac{\alpha}{1-\Lambda^{-\frac{1}{2}}}.
\end{align}
For $i\gg m$ it decays as $\Lambda^{-(i-m)/2}$, while for $i\ll m$ it decays as $\Lambda^{-(m-i)/2}$.

We then calculate the bath block of the single-particle density matrix by summing over the occupied negative-energy eigenstates:
\begin{align}
    \rho_{i\eta,j\eta'}
    =\sum_m u^*_{i\eta,m-}u_{j\eta',m-}
    &=\alpha^2\sum_{m=1}^{\infty}
    \frac{\Lambda^{\frac{i+j-2m}{2}}}
    {(-\Lambda^{i-m-\frac{1}{2}}-\eta)
     (-\Lambda^{j-m-\frac{1}{2}}-\eta')}
    \nonumber\\
    &=\alpha^2\sum_{r=-\infty}^{\infty}
    \frac{\Lambda^{\frac{i-j+2r}{2}}}
    {(-\Lambda^{i-j+r-\frac{1}{2}}-\eta)
     (-\Lambda^{r-\frac{1}{2}}-\eta')}
    \nonumber\\
    &=\alpha^2\Lambda^{\frac{i-j}{2}}
    \sum_{r=-\infty}^{\infty}
    \frac{\Lambda^r}
    {(\Lambda^{i-j+r-\frac{1}{2}}+\eta)
     (\Lambda^{r-\frac{1}{2}}+\eta')}\, .
\end{align}
Here the second line uses $r=j-m$.

We define
\begin{align}
    \lambda_-^2
    =\sum_{r=-\infty}^\infty
    \frac{\Lambda^r}{(\Lambda^{r-\frac{1}{2}}-1)^2},\quad
    \lambda_+^2=\sum_{r=-\infty}^\infty
    \frac{\Lambda^r}{(\Lambda^{r-\frac{1}{2}}+1)^2},
\end{align}
so that $\alpha^2(\lambda_+^2+\lambda_-^2)=1$.

\paragraph{$\Lambda\to\infty$ limit.}
We now consider the large $\Lambda$ limit.
\begin{align}
    \lambda_-^2
    =\sum_{r=-\infty}^\infty
    \frac{\Lambda^r}{(\Lambda^{r-\frac{1}{2}}-1)^2}
    =\sum_{r=0}^{\infty}
    \frac{2\Lambda^{-r}}{(1-\Lambda^{-r-\frac{1}{2}})^2}
    &=\sum_{r=0}^{\infty}\sum_{a=0}^{\infty}
    2(a+1)\Lambda^{-[r+(r+\frac{1}{2})a]}
    \nonumber\\
    &= 2(1+2\Lambda^{-1/2}+4\Lambda^{-1}+4\Lambda^{-3/2}+6\Lambda^{-2}+...) \\
    \lambda_+^2
    =\sum_{r=-\infty}^\infty
    \frac{\Lambda^r}{(\Lambda^{r-\frac{1}{2}}+1)^2}
    =\sum_{r=0}^{\infty}
    \frac{2\Lambda^{-r}}{(1+\Lambda^{-r-\frac{1}{2}})^2}
    &=\sum_{r=0}^{\infty}\sum_{a=0}^{\infty}
    2(-1)^a(a+1)\Lambda^{-[r+(r+\frac{1}{2})a]}
    \nonumber\\
    &= 2(1-2\Lambda^{-1/2}+4\Lambda^{-1}-4\Lambda^{-3/2}+6\Lambda^{-2}-...)\\
    \alpha^{-2}
    =\lambda_+^2+\lambda_-^2
    =\sum_{r=0}^{\infty}\sum_{a=0}^{\infty}
    4(2a+1)\Lambda^{-[r+(r+\frac{1}{2})2a]}
    &=4(1+4\Lambda^{-1}+6\Lambda^{-2}+...)
\end{align}
Defining the branch occupations by $n_\eta\equiv\rho_{i\eta,i\eta}=\alpha^2\lambda_\eta^2$, we obtain
\begin{align}
    \alpha
    &=\frac{1}{2}-\Lambda^{-1}
    +\frac{3}{2}\Lambda^{-2}+O(\Lambda^{-3}),\nonumber\\
    n_-
    &=\frac{1}{2}+\Lambda^{-\frac{1}{2}}
    -2\Lambda^{-\frac{3}{2}}
    +O(\Lambda^{-\frac{5}{2}}),\qquad
    n_+=1-n_- .
\end{align}

The off-diagonal element for $i>j$ is expanded as
\begin{align}
    \rho_{i\eta,j\eta'}
    &=\alpha^2\Lambda^{\frac{i-j}{2}}
    \sum_{r=-\infty}^{\infty}
    \frac{\Lambda^r}
    {(\Lambda^{i-j+r-\frac{1}{2}}+\eta)
     (\Lambda^{r-\frac{1}{2}}+\eta')}
    \nonumber\\
    &=\alpha^2\Lambda^{\frac{i-j}{2}}
    \left(
    \sum_{r=-\infty}^{j-i}
    +\sum_{r=j-i+1}^{0}
    +\sum_{r=1}^{\infty}
    \right)
    \frac{\Lambda^r}
    {(\Lambda^{i-j+r-\frac{1}{2}}+\eta)
     (\Lambda^{r-\frac{1}{2}}+\eta')}
    \\
    &=\alpha^2\Lambda^{\frac{i-j}{2}}
    \Bigg[
    \sum_{r=0}^{\infty}
    \frac{\Lambda^{j-i-r}}
    {(\Lambda^{-r-\frac{1}{2}}+\eta)
     (\Lambda^{j-i-r-\frac{1}{2}}+\eta')}
    \nonumber\\
    &\qquad
    +\sum_{r=1}^{i-j}
    \frac{\Lambda^{j-i+r}}
    {(\Lambda^{r-\frac{1}{2}}+\eta)
     (\Lambda^{j-i+r-\frac{1}{2}}+\eta')}
    +\sum_{r=1}^{\infty}
    \frac{\Lambda^r}
    {(\Lambda^{i-j+r-\frac{1}{2}}+\eta)
     (\Lambda^{r-\frac{1}{2}}+\eta')}
    \Bigg].
\end{align}
To lowest order, the three terms are $ \alpha^2\eta\eta' \Lambda^{-\frac{i-j}{2}},\alpha^2\eta'\Lambda^{-\frac{i-j-1}{2}},\alpha^2\Lambda^{-\frac{i-j}{2}}$.

To zeroth order in $\Lambda^{-1/2}$, the two diagonal branch occupations are $n_+=n_-=\frac{1}{2}$ and the nonzero nearest-shell off-diagonal elements have magnitude $1/4$. The single-particle density matrix is
\begin{align}
    \begin{pmatrix}
        \frac{1}{2} & 0 &\frac{1}{4} & \frac{1}{4} &0 &0 &\cdots\\
        0 &\frac{1}{2} & -\frac{1}{4} & -\frac{1}{4} &0 &0  &\cdots \\
        \frac{1}{4} & -\frac{1}{4} & \frac{1}{2} & 0 &\frac{1}{4} & \frac{1}{4} &\cdots \\
        \frac{1}{4} & -\frac{1}{4}  & 0& \frac{1}{2} &-\frac{1}{4} & -\frac{1}{4} &\cdots \\
        0 & 0 &\frac{1}{4} & -\frac{1}{4} & \frac{1}{2} & 0  &\cdots \\
        0& 0& \frac{1}{4} & -\frac{1}{4}  & 0& \frac{1}{2} &\cdots
    \end{pmatrix} \label{eq:rdm-largeLam-0th}
\end{align}

We can apply the unitary transformation $\frac{1}{\sqrt{2}}\begin{pmatrix}
    1 & 1\\
    1 & -1
\end{pmatrix}$ to each block, obtaining
\begin{align}
    \begin{pmatrix}
        \frac{1}{2} & 0 & 0 & 0 &0 &0 &\cdots\\
        0 &\frac{1}{2} & \frac{1}{2} & 0 &0 &0  &\cdots \\
        0 & \frac{1}{2} & \frac{1}{2} & 0 & 0 & 0 &\cdots \\
        0 & 0 & 0& \frac{1}{2} & \frac{1}{2} & 0 &\cdots \\
        0 & 0 &0 & \frac{1}{2} & \frac{1}{2} & 0  &\cdots \\
        0& 0& 0 & 0  & 0& \frac{1}{2} &\cdots
    \end{pmatrix}
\end{align}
All eigenvalues are either $0$ or $1$, except for one eigenvalue equal to $\frac{1}{2}$, and the entanglement entropy is $\ln2$. This statement holds for the semi-infinite matrix, which has only one boundary. For a finite truncation, another $\frac{1}{2}$ eigenvalue appears at the artificial far boundary. Numerically, we find that when $\Lambda\to \infty$ $S\to \ln2$, which is consistent with this result.

To obtain the next-order contribution to the entanglement entropy, we expand $u$ in $\Lambda^{-1/2}$. Since $|u_{i\eta,m\zeta}|$ decays quickly with $|i-m|$, eigenstates centred far from the entanglement cut are either fully occupied or fully empty in the retained subsystem and do not contribute to the entropy. Hence, only finitely many eigenstates and bath sites near the cut are needed. The first nonzero correction occurs at order $\Lambda^{-2}$. For a fixed eigenstate $(m,\zeta)$, the required components are
\begin{align}
    u_{m\eta,m\zeta}
    &=\frac{\alpha}{\zeta\Lambda^{-\frac{1}{2}}-\eta}=-\alpha\left(
    \eta+\zeta\Lambda^{-\frac{1}{2}}
    +\eta\Lambda^{-1}
    +\zeta\Lambda^{-\frac{3}{2}}
    +\eta\Lambda^{-2}+\cdots\right),\\
    u_{(m+1)\eta,m\zeta}
    &=\frac{\alpha}{\zeta-\eta\Lambda^{-\frac{1}{2}}}=\alpha\left(
    \zeta+\eta\Lambda^{-\frac{1}{2}}
    +\zeta\Lambda^{-1}
    +\eta\Lambda^{-\frac{3}{2}}
    +\zeta\Lambda^{-2}+\cdots\right),\\
    u_{(m-1)\eta,m\zeta}
    &=\frac{\alpha\Lambda^{-\frac{1}{2}}}
    {\zeta\Lambda^{-\frac{3}{2}}-\eta}
    =-\alpha\Lambda^{-\frac{1}{2}}
    \left(\eta+\zeta\Lambda^{-\frac{3}{2}}+\cdots\right),\\
    u_{(m+2)\eta,m\zeta}
    &=\frac{\alpha\Lambda}
    {\zeta\Lambda^{\frac{3}{2}}-\eta}
    =\alpha\Lambda^{-\frac{1}{2}}
    \left(\zeta+\eta\Lambda^{-\frac{3}{2}}+\cdots\right),\\
    u_{(m-2)\eta,m\zeta}
    &=\frac{\alpha\Lambda^{-1}}
    {\zeta\Lambda^{-\frac{5}{2}}-\eta}
    =-\alpha\eta\Lambda^{-1},\\
    u_{(m+3)\eta,m\zeta}
    &=\frac{\alpha\Lambda^{\frac{3}{2}}}
    {\zeta\Lambda^{\frac{5}{2}}-\eta}
    =\alpha\zeta\Lambda^{-1},\\
    u_{(m-3)\eta,m\zeta}
    &=\frac{\alpha\Lambda^{-\frac{3}{2}}}
    {\zeta\Lambda^{-\frac{7}{2}}-\eta}
    =-\alpha\eta\Lambda^{-\frac{3}{2}},\\
    u_{(m+4)\eta,m\zeta}
    &=\frac{\alpha\Lambda^2}
    {\zeta\Lambda^{\frac{7}{2}}-\eta}
    =\alpha\zeta\Lambda^{-\frac{3}{2}},\\
    u_{(m-4)\eta,m\zeta}
    &=\frac{\alpha\Lambda^{-2}}
    {\zeta\Lambda^{-\frac{9}{2}}-\eta}
    =-\alpha\eta\Lambda^{-2},\\
    u_{(m+5)\eta,m\zeta}
    &=\frac{\alpha\Lambda^{\frac{5}{2}}}
    {\zeta\Lambda^{\frac{9}{2}}-\eta}
    =\alpha\zeta\Lambda^{-2}.
\end{align}
Recall that $\alpha = \frac{1}{2} - \Lambda^{-1} + \frac{3}{2}\Lambda^{-2} + O(\Lambda^{-3})$. To zeroth order in $\Lambda^{-1/2}$, the single-particle density matrix near a cut at shell $i$ is obtained from $\sum_{m=i-1}^{i+1}\mathbf{u}_{m-}\mathbf{u}_{m-}^\dagger$, which reproduces the $6\times6$ subblocks in \cref{eq:rdm-largeLam-0th}. At the next nonvanishing order, it is sufficient to retain $\sum_{m=i-4}^{i+5}\mathbf{u}_{m-}\mathbf{u}_{m-}^\dagger$, whose nonzero matrix elements remain localized near the cut. Using Mathematica, we verify that the leading correction to the eigenvalues of the restricted single-particle density matrix is of order $\Lambda^{-2}$. The eigenvalues with the largest deviations from $0$ and $1$ are $\frac{1}{4}\Lambda^{-2}-2\Lambda^{-3}$ and $1-\frac{1}{4}\Lambda^{-2}+2\Lambda^{-3}$, while the remaining deviations occur at higher orders in $\Lambda^{-1/2}$. Therefore, the leading logarithmic correction to the entanglement entropy is $-2\cdot\frac{1}{4}\Lambda^{-2}\cdot \ln (\frac{1}{4}\Lambda^{-2})$. This is also confirmed numerically, as shown in \cref{fig:topo}(a).

The bulk single-particle density matrix can also be understood through its Fourier symbol:
\begin{align}
    \td{\rho}_\infty(q) &=  \begin{pmatrix}
        \frac{1}{2}+\frac{1}{2}\cos q & -\frac{\ii}{2}\sin q \\
       \frac{\ii}{2}\sin q &\frac{1}{2}-\frac{1}{2}\cos q
    \end{pmatrix}  = \frac{1}{2}\tau_0 + \frac{1}{2}(\tau_z\cos q+\tau_y\sin q)
\end{align}
which has winding number one and enforces a $\frac{1}{2}$ eigenvalue.

\paragraph{$\Lambda\to1$ limit.} We then consider the opposite limit, $\Lambda\to1$, taking the bulk limit described above before sending $\Lambda$ to one.
Let $\Lambda=e^{\delta_\Lambda}$, $x=r\delta_\Lambda$, and $y=(j-i)\delta_\Lambda$. Away from the diagonal singularity, we then have
\begin{align}
    \rho_{i\eta,j\eta'} = \alpha^2 e^{-\frac{y}{2}}\cdot \frac{1}{\delta_\Lambda}\cdot \int_{-\infty}^{\infty}{\rm d}x  \frac{e^x}{(e^{x-y}+\eta)(e^{x}+\eta')}\, .
\end{align}

We then calculate
\begin{align}
    I_{\eta,\eta'}(y) &= \int_{-\infty}^{\infty}{\rm d}x  \frac{e^x}{(e^{x-y}+\eta)(e^{x}+\eta')}= \int_{0}^{\infty} {\rm d}t \frac{1}{(te^{-y}+\eta)(t+\eta')} = \frac{1}{\eta' e^{-y}-\eta}\ln\Bigg|\frac{\eta' e^{-y}}{\eta}\Bigg| = \frac{ye^y}{\eta e^y-\eta'},
    \qquad t=e^x .
\end{align}
where we take principal values when the integrand has a pole on the real axis. However, it is not correct for $y=0$ and $\eta=\eta'=-1$, where a divergent second-order pole $\frac{1}{(t-1)^2}$ lies within the integration domain. The discrete sum imposes cutoffs around $t=e^{\pm\frac{1}{2}\delta_\Lambda}$, so $I_{-,-}(0)\propto\frac{1}{\delta_\Lambda}$. To calculate this divergent behavior, we return to the original summation
\begin{align}
    \rho(0)_{--}
    &=
    \alpha^2\sum_{r=-\infty}^{\infty}
    \frac{\Lambda^r}{(\Lambda^{r-\frac{1}{2}}-1)^2}
    =
    \alpha^2\sum_{r=-\infty}^{\infty}
    \frac{e^{r\delta_\Lambda}}{(e^{(r-\frac{1}{2})\delta_\Lambda}-1)^2}
    \nonumber\\
    &=
    \frac{\alpha^2}{\delta_\Lambda^2}
    \sum_{r=-\infty}^{\infty}
    \frac{1}{(r-\frac{1}{2})^2}
    +O\left(\frac{1}{\delta_\Lambda}\right)
    =
    \frac{\alpha^2\pi^2}{\delta_\Lambda^2}
    +O\left(\frac{1}{\delta_\Lambda}\right).
\end{align}
The $O(1/\delta_\Lambda)$ contribution cannot be calculated by expanding each summand in $\delta_\Lambda$. The summation diverges as $|r|\to\infty$ unless the regularizing factor $e^{-r\delta_\Lambda}$ is retained, so this factor cannot be expanded term by term. Guided by the integral asymptotics and the normalization of the projector, we use the bulk parametrization $\rho(0)_{++}=\frac{\delta_\Lambda}{\pi^2}$ and $\rho(0)_{--}=1-\frac{\delta_\Lambda}{\pi^2}$. Then the smooth part is
\begin{align}
    \rho_{i\eta,j\eta'} = \frac{\delta_\Lambda}{\pi^2}\frac{ye^{\frac{y}{2}}}{\eta e^y-\eta'}
\end{align}
This normalization is also verified by numerical calculation.

To obtain the eigenvalues of $\rho_{i\eta,j\eta'}$, we replace the shell summation by an integral. Suppressing the branch indices, the matrix eigenvalue equation is
\begin{align}
    \sum_j\rho_{ij}\phi_j=\kappa\phi_i ,
\end{align}
where $\kappa$ is an eigenvalue of the single-particle density matrix. Writing $x=i\delta_\Lambda$ and $y=(j-i)\delta_\Lambda$, the continuum limit becomes
\begin{align}
    \int{\rm d}y\,
    \varrho(y)\phi(x+y)=\kappa\phi(x),
\end{align}
where the continuum single-particle density kernel is defined by the distributional scaling limit
\begin{align}
    \varrho_{\eta,\eta'}(y)
    \equiv
    \lim_{\delta_\Lambda\to0}
    \frac{\rho_{i\eta,j\eta'}}{\delta_\Lambda}
    =
    \frac{1}{\pi^2}\frac{ye^{\frac{y}{2}}}{\eta e^y-\eta'}
    +\delta(y)\delta_{\eta,-}\delta_{\eta',-}.
\end{align}
Here $y=(j-i)\delta_\Lambda$ is held fixed in the limit. The $\delta$-function comes from the $O(1)$ diagonal element $\rho(0)_{--}$. One can check that its prefactor is correct by comparing $\sum_j\rho_{ij}$ with $\int{\rm d}y\,\varrho(y)$. Translation invariance implies that the coordinate dependence of an eigenfunction is $e^{\ii px}$. The corresponding eigenvalues are obtained by diagonalizing
\begin{align}
    \td{\varrho}_{\eta\eta'}(p)
    =
    \int{\rm d}y\,
    \varrho_{\eta\eta'}(y)e^{\ii py},
\end{align}
where $p=q/\delta_\Lambda=q/\ln\Lambda$ is the continuum Fourier momentum. For $q\in[-\pi,\pi)$, its range approaches $(-\infty,\infty)$ as $\Lambda\to1$. Equivalently, the continuum and discrete Fourier symbols are related by
\begin{align}
    \td{\varrho}(p)
    =
    \lim_{\Lambda\to1}
    \td{\rho}_\Lambda(\delta_\Lambda p).
\end{align}
Explicitly,
\begin{align}
    \td{\varrho}(p) &= \begin{pmatrix}
        \frac{1}{\cosh^2 \pi p} & -\ii \frac{\tanh \pi p}{\cosh \pi p} \\
        \ii \frac{\tanh \pi p}{\cosh \pi p} & 1-\frac{1}{\cosh^2 \pi p}
    \end{pmatrix} \nonumber \\
    & = \frac{1}{\cosh^2 \pi p}\begin{pmatrix}
        1 & -\ii \sinh(\pi p) \\
        \ii \sinh(\pi p) & \sinh^2(\pi p)
    \end{pmatrix} \nonumber \\
    & = \frac{1}{2}\tau_0 + \frac{1-\sinh^2\pi p}{2\cosh^2\pi p}\tau_z + \frac{\sinh\pi p}{\cosh^2\pi p}\tau_y
\end{align}
The eigenvalues are
\begin{align}
    &\frac{1}{2}\pm  \frac{1}{2\cosh^2\pi p}\sqrt{(1-\sinh^2\pi p)^2+4\sinh^2\pi p} = \frac{1}{2}\pm \frac{1}{2}= 0,1
\end{align}
and the eigenstates are
\begin{align}
    \begin{pmatrix}
        \ii \tanh(\pi p)\\
        {\rm sech}(\pi p)
    \end{pmatrix}, \begin{pmatrix}
        {\rm sech}(\pi p) \\
        \ii \tanh(\pi p)
    \end{pmatrix}
\end{align}
Let $\td{\varrho}(p)=\frac{1}{2}\tau_0+\td{\varrho}_z(p)\tau_z+\td{\varrho}_y(p)\tau_y$.
The map $p\mapsto(\td{\varrho}_z,\td{\varrho}_y)$ covers the circle of radius $\frac{1}{2}$. We can identify $p=\infty$ and $p=-\infty$ because
\begin{align}
    \lim_{p\to\infty}\td{\varrho}(p)
    =
    \lim_{p\to-\infty}\td{\varrho}(p)
    =
    \frac{1}{2}\tau_0-\frac{1}{2}\tau_z .
\end{align}
The map therefore also has winding number one.

The eigenvalues of $\td{\varrho}(p)$ are all $0$ or $1$, corresponding to a pure state of fixed point wavefunction. We then truncate the system and compute the entanglement entropy on the half-line. Numerically, the truncated density kernel has an edge eigenvalue $1/2$, protected by the nontrivial topology. The remaining eigenvalues deviate from $0$ or $1$ as $\Delta n_r\approx A\gamma^{-r}$, $r=0,1,2,\ldots$, where $A\approx4.6\times10^{-3}$ and $\gamma>1$ is the fitted decay factor, producing the finite-$\Lambda$ corrections.

\subsection{Kondo model}
For the single-channel spin-$\frac{1}{2}$ Kondo model, the fixed point is at $J_K\to\infty$. Therefore, the singlet $\frac{1}{\sqrt{2}}(c^\dagger_\down(0)|\up\rangle_f - c_\up^\dagger(0)|\down\rangle_f)$ is formed. To compare with the Anderson impurity model, it is useful to work first in the Wilson chain representation. At the strong-coupling fixed point, the first Wilson site is locked into a singlet with the impurity, so the full ground state factorizes into this local singlet and the ground state of the remaining Wilson chain. This is the same fixed-point structure as the $t\to\infty$ limit of the $U=0$ Anderson model. The transformation from the Wilson chain to the star representation is a bath-only unitary fixed by the discretized bath and is identical for the two models. Therefore the fixed-point correlation matrix in the star basis is also identical, and the fixed-point energy-space EE is exactly the same for any value of $\Lambda$ in the single-channel spin-$\frac{1}{2}$ case, as shown in \cref{fig:FL}(b) and \cref{fig:topo}(a). The same Fermi-liquid fixed-point picture is used for fully screened cases with larger $S_{\rm imp}$, whose fixed-point entropies are supported numerically in the main text.

\subsection{Analysis of symmetry and topology \label{app:sym-topo}}
In this section we analyze how Altland-Zirnbauer symmetries constrain both the impurity and the energy-space bath, and how these constraints determine the topology of the free-fermion fixed point.

\paragraph{Altland-Zirnbauer class.}
The Altland-Zirnbauer classification of free-fermion systems is based on the presence or absence of time-reversal symmetry $\mathcal{T}$, particle-hole symmetry $\mathcal{C}$, and chiral symmetry $\mathcal{S}$. We first state the symmetry action on the impurity, derive the corresponding constraints on the local impurity Hamiltonian and the spectral hybridization, and then construct an effective energy-space Hamiltonian.

Integrating out the bath, we obtain the effective action of the impurity in imaginary time as
\begin{align}
    S_{eff} = \sum_{\ii\omega_n}\sum_{\alpha\beta} f^\dagger_\alpha(\ii\omega_n)\left(\ii\omega_n \delta_{\alpha\beta} - [H_f]_{\alpha\beta} - \td{\Delta}_{\alpha\beta}(\ii\omega_n) \right) f_\beta(\ii\omega_n)
\end{align}
Here $H_f$ denotes the first-quantized local impurity Hamiltonian matrix of the effective quadratic model; its second-quantized contribution is $\sum_{\alpha\beta}[H_f]_{\alpha\beta}f^\dagger_\alpha f_\beta$. It is distinct from the many-body impurity Hamiltonian $\hH_f$ used earlier, which may contain interactions. We use $\alpha,\beta$ as generic flavor indices that include all spin and orbital labels, $\alpha,\beta=1,\ldots,N_f$. $\td{\Delta}(z)$ is the hybridization function in the complex-frequency domain, which is a matrix in flavor space and is related to the spectral hybridization function by
\begin{align}
    \Delta(\omega)
    =\frac{\ii}{2\pi}\left[
    \td{\Delta}(\omega+\ii0^+)
    -\td{\Delta}(\omega-\ii0^+)\right].
\end{align}

The symmetries $\mathcal{T},\mathcal{C},\mathcal{S}$ act on the impurity fermion as
\begin{align}
    &\mathcal{T} f_\alpha \mathcal{T}^{-1} = \sum_{\beta} (U_T^{f})^*_{\beta\alpha} f_\beta,\ \mathcal{T}\ii \mathcal{T}^{-1} = -\ii \\
    &\mathcal{C} f_\alpha\mathcal{C}^{-1} = \sum_{\beta} (U_{C}^{f})_{\beta\alpha} f^\dagger_\beta,\ \mathcal{C}\ii \mathcal{C}^{-1} = \ii \\
    &\mathcal{S} f_\alpha\mathcal{S}^{-1} = \sum_{\beta} (U_S^{f})_{\beta\alpha} f^\dagger_\beta,\ \mathcal{S}\ii \mathcal{S}^{-1} = -\ii
\end{align}
where $U_T^{f},U_C^{f},U_S^{f}$ are unitary matrices, and the operators $\mathcal{T},\mathcal{C},\mathcal{S}$ always commute with the many-body Hamiltonian if the symmetries are preserved. Here $\mathcal{T}$, $\mathcal{C}$, and $\mathcal{S}$ denote their many-body implementations on Fock space. In this convention, $\mathcal{T}$ and $\mathcal{S}$ are antiunitary, whereas $\mathcal{C}$ is unitary; the corresponding first-quantized representatives are $U_TK$, $U_CK$, and the unitary chiral operator $U_S$, respectively. Using $f_{\alpha}(z) = \int {\rm d }\tau e^{z\tau} e^{\tau \hH}f_\alpha e^{-\tau \hH},f_{\alpha}^\dagger(z) = \int {\rm d }\tau e^{-z\tau} e^{\tau \hH}f^\dagger_\alpha e^{-\tau \hH} $, we then have
\begin{align}
    &\mathcal{T} f_\alpha(z) \mathcal{T}^{-1} = \sum_{\beta} (U_T^{f})^*_{\beta\alpha} f_\beta(z^*),\\
    &\mathcal{C} f_\alpha(z) \mathcal{C}^{-1} = \sum_{\beta} (U_{C}^f)_{\beta\alpha} f_\beta^\dagger(-z), \\
    &\mathcal{S} f_\alpha(z) \mathcal{S}^{-1} = \sum_{\beta} (U_S^{f})_{\beta\alpha} f^\dagger_\beta(-z^*),
	\end{align}
	in the frequency domain.

	The local impurity single-particle Hamiltonian is constrained by the same symmetry action. In the above convention,
	\begin{align}
	    (U_T^{f})^\dagger H_f U_T^{f} &= (H_f)^*,&
	    (U_C^{f})^\dagger H_f U_C^{f} &= -(H_f)^*,&
	    (U_S^{f})^\dagger H_f U_S^{f} &= -H_f .
	    \label{eq:Hf-symmetry-constraints}
	\end{align}
	The bath enters the impurity problem through the hybridization function. Requiring $\sum_{z,\alpha\beta} f_\alpha^\dagger(z)\td{\Delta}_{\alpha\beta}(z)f_\beta(z)$ to be invariant, we obtain the symmetry constraints on $\td{\Delta}(z)$ as
	\begin{align}
	    (U_T^{f})^\dagger\td{\Delta}(z) U_T^{f} = \td{\Delta}^*(z^*),\ (U_C^{f})^\dagger\td{\Delta}(z) U_C^{f} = -\td{\Delta}^T(-z),\ (U_S^{f})^\dagger\td{\Delta}(z) U_S^{f} = -\td{\Delta}^\dagger(-z^*)
	\end{align}
	$\Delta(\omega)$ should then satisfy
	\begin{align}
	    (U_T^{f})^\dagger\Delta(\omega) U_T^{f} = \Delta^*(\omega),\ (U_C^{f})^\dagger\Delta(\omega) U_C^{f} = \Delta^*(-\omega),\ (U_S^{f})^\dagger\Delta(\omega) U_S^{f} = \Delta(-\omega)
	\end{align}
	Together with \cref{eq:Hf-symmetry-constraints}, these constraints show that the AZ class of the full quadratic fixed-point Hamiltonian is determined by the symmetries preserved simultaneously by $H_f$ and by the bath hybridization $\Delta(\omega)$. This gives two equivalent ways to realize a target class. One may keep the bath more symmetric and use an allowed impurity term $H_f$ to break the accidental symmetries down to the desired class, or keep the impurity more symmetric and instead break the bath hybridization to the desired class. In both cases the protected low-energy entropy is controlled by the symmetry class of the complete energy-space Hamiltonian, not by whether the symmetry reduction is implemented on the impurity or bath side.

	For a hybridization function satisfying the above symmetry constraints, we can define an effective energy-space Hamiltonian that obeys the same symmetries, although the symmetry action on the effective bath is not unique.  For a generic multiorbital Anderson impurity model, this Hamiltonian can be written as
\begin{align}
    \hH = \sum_{\alpha\beta}[H_f]_{\alpha\beta}f^\dagger_\alpha f_\beta
    +\sum_{\alpha\beta}\sum_{\eta=\pm}\sum_{i=1}^L ( [t_{i\eta}]_{\alpha\beta}  f^\dagger_{\alpha} c_{i\eta \beta} +h.c.) +  \sum_{\alpha}\sum_{\eta=\pm}\sum_{i=1}^L \eta \varepsilon_i  c^\dagger_{i\eta \alpha} c_{i\eta \alpha} \label{eq:HAIM-dis-multiorb}
\end{align}
where $t_{i\eta}t_{i\eta}^\dagger=\int_{i,\eta} {\rm d}\varepsilon \Delta(\varepsilon)\equiv\Delta_{i\eta}$ and $\int_{i,\eta} {\rm d}\varepsilon$ is the integral over the energy range of the $i$-th bath level on the $\eta$ branch, as defined in \cref{app:discretization}. A right-unitary gauge transformation $t_{i\eta}\to t_{i\eta}V_{i\eta}$, with $V_{i\eta}$ unitary and accompanied by the corresponding rotation of the bath basis, leaves $t_{i\eta}t_{i\eta}^\dagger$ unchanged and does not change the entanglement between different energy scales.
Here we explicitly write down a choice of $t_{i\eta}$ and the corresponding action of symmetries. For a physical fermionic bath, $\td{\Delta}(z)$ has a spectral representation $\td{\Delta}(z)=\int {\rm d}\omega\,\Delta(\omega)/(z-\omega)$, with $\Delta_{\alpha\beta}(\omega)=\sum_a v_{\alpha a}(\omega)v^*_{\beta a}(\omega)$. Therefore $\Delta(\omega)$ is Hermitian and positive semidefinite for each real $\omega$. We can diagonalize $\Delta_{i\eta} = U_{i\eta}T_{i\eta}U^\dagger_{i\eta}$ where $T_{i\eta}$ is a diagonal matrix with real and non-negative diagonal elements, and $t_{i\eta} =  U_{i\eta}\sqrt{T_{i\eta}}$. To uniquely determine $t_{i\eta}$, we need to fix the gauge of $U_{i\eta}$. The symmetries $\mathcal{T},\mathcal{C},\mathcal{S}$ constrain the first-quantized Hamiltonian $H$ to satisfy $U_T H^* U_T^\dag = H,U_C H^* U_C^\dag = -H,U_S H U_S^\dag = -H$, respectively. The constraints on $t$ are
\begin{align}
    U_T^f t_{i\eta}^* &= \sum_{\eta'}t_{i\eta'} [U_T^c]_{i\eta',i\eta}\ , \\
    U_C^f t_{i\eta}^* &= -\sum_{\eta'}t_{i\eta'} [U_C^c]_{i\eta',i\eta}\ , \\
    U_S^f t_{i\eta} &= -\sum_{\eta'}t_{i\eta'} [U_S^c]_{i\eta',i\eta}\ ,
\end{align}
respectively, where $U_T^c,U_C^c,U_S^c$ are the representations of the symmetries on the bath degrees of freedom, $[...]_{i\eta,i\eta'}$ denotes the submatrix corresponding to the indices $i\eta$ and $i\eta'$, and $\bar\eta\equiv-\eta$ denotes the opposite energy branch.
We then discuss the choice of $U_{i\eta}$ and symmetries one by one.

\begin{itemize}
    \item $\mathcal{T}^2=1$. In this case, the eigenvectors of $\Delta_{i\eta}$ are eigenstates of $U^f_TK$, where $K$ is the complex conjugate. Therefore, $U^f_T U^*_{i\eta} = U_{i\eta}$ and $\hH$ respects $\mathcal{T}$ when $U_T^c$ is an identity matrix.
    \item $\mathcal{T}^2=-1$. In this case, $N_f$ should be even due to the Kramers degeneracy, and the eigenvectors of $\Delta_{i\eta}$ are Kramers pairs. They satisfy $U^f_T U^*_{i\eta} = U_{i\eta} \begin{pmatrix}
        \ii\sigma_y & 0 &\cdots \\
        0 & \ii\sigma_y &\cdots \\
        \vdots & \vdots  & \ddots
    \end{pmatrix}$. $\hH$ respects $\mathcal{T}$ when $[U_T^c]_{i\eta\alpha,i'\eta'\alpha'} = \delta_{ii'}\delta_{\eta\eta'}\begin{pmatrix}
        \ii\sigma_y & 0 &\cdots \\
        0 & \ii\sigma_y &\cdots \\
        \vdots & \vdots  & \ddots
    \end{pmatrix}_{\alpha\alpha'}$.
    \item $\mathcal{C}^2=1$. In contrast to $\mathcal{T}$, which does not change the $\eta$ index, $\mathcal{C}$ should flip the $\eta$ index so that $U_C$ anticommutes with the on-site bath terms. Therefore, we choose $[U^c_C]_{i\eta\alpha,i'\eta'\alpha'} = \delta_{ii'}(\tau_x)_{\eta\eta'}\delta_{\alpha\alpha'}=\delta_{ii'}\delta_{\eta,\bar{\eta}'}\delta_{\alpha\alpha'}$. Since $\Delta_{i\bar\eta}=U^f_C\Delta^*_{i\eta}U^{f\dagger}_C$, $U_C^fU^*_{i\eta}$ is the eigenvector matrix of $\Delta_{i\bar\eta}$, with $T_{i\bar\eta}=T_{i\eta}$. Choosing $U_{i\bar\eta} = -U^f_C U^*_{i\eta}$, we have $U_CH^*U_C^\dagger=-H$.
    \item $\mathcal{C}^2=-1$. We choose $[U^c_C]_{i\eta\alpha,i'\eta'\alpha'} = \delta_{ii'}(\ii\tau_y)_{\eta\eta'}\delta_{\alpha\alpha'}=\eta \delta_{ii'}\delta_{\eta,\bar{\eta}'}\delta_{\alpha\alpha'}$. When $U_{i\bar\eta} = \eta U^f_C U^*_{i\eta}$, we have $U_CH^*U_C^\dagger=-H$.
    \item $\mathcal{S}$. Anticommutation between $U_S$ and the on-site bath terms also requires $U_S$ to flip the $\eta$ index, and we choose $[U^c_S]_{i\eta\alpha,i'\eta'\alpha'} = \delta_{ii'}(\tau_x)_{\eta\eta'}\delta_{\alpha\alpha'}=\delta_{ii'}\delta_{\eta,\bar{\eta}'}\delta_{\alpha\alpha'}$. Since $U^f_S \Delta_{i\eta} U^{f\dagger}_S = \Delta_{i\bar\eta}$, $U_S^f U_{i\eta}$ is the eigenvector matrix of $\Delta_{i\bar\eta}$, with $T_{i\bar\eta}=T_{i\eta}$. Choosing $U_{i\bar\eta} = -U^f_S U_{i\eta}$, we have $U_SHU_S^\dagger=-H$.
    \item When both $\mathcal{T}$ and $\mathcal{C}$ exist, we can choose $U_T^c,U_C^c$ as above and then $U_S^c = U_T^c (U_C^c)^*$.
	\end{itemize}

	\Cref{tab:az-local-hf} lists the impurity-space matrices $U_T^f$, $U_C^f$, and $U_S^f$, denoted below by $U_T$, $U_C$, and $U_S$. We represent a symmetry squaring to $+1$ by $\mathbb{1}_{N_f}$. A symmetry squaring to $-1$ requires even $N_f$, and we represent it by the standard antisymmetric matrix $J_{N_f}\equiv\mathbb{1}_{N_f/2}\otimes\ii\sigma_y$. Here $\sigma_y$ acts on the two-dimensional factor and $\mathbb{1}_{N_f/2}$ on the remaining orbital/flavor indices.

	For $\mathcal{T}^2=-1$, $\sigma_y$ acts on the physical-spin (Kramers) doublet. This interpretation applies to AII and to the time-reversal sectors of DIII and CII. For $\mathcal{C}^2=-1$, the interpretation also depends on the basis. In a non-BdG annihilation-operator basis, $\sigma_y$ acts on paired internal-state indices, while $\mathcal C$ maps annihilation operators to creation operators. For C and CI, we may instead choose a two-component BdG basis, in which $\sigma_y$ acts on a particle--hole doublet formed by a particle and a hole with opposite spins. The DIII representative uses $U_C=\mathbb{1}_{N_f}$; D and BDI also use $U_C=\mathbb{1}_{N_f}$ and introduce no $\mathcal C$-associated $J_{N_f}$. At the minimal dimension for CII, $N_f=2$ and $U_T=U_C=J_2=\ii\sigma_y$. This two-dimensional matrix representation applies only to a non-BdG basis. An explicit BdG representation of CII must include both spin/Kramers and particle--hole doublets and therefore requires at least a four-component basis.

	\begin{table*}[t]
	    \caption{Symmetry representations and direct local impurity terms used for the constant-hybridization representative. A dash denotes an absent symmetry. Here $N_f$ is unrestricted for A, AIII, AI, BDI, and D, and even for DIII, AII, CII, C, and CI. In the tensor-product expressions in the last column, $\sigma_\mu$ acts on the physical-spin (Kramers) factor for DIII and AII. For C and CI, it acts on a paired internal-state factor in a non-BdG basis or on the particle--hole factor in a two-component BdG basis formed by a particle and a hole with opposite spins. The first factor acts on the remaining $N_f/2$-dimensional orbital/flavor space. The same convention applies to the corresponding hybridization matrices below. For the plotted data, we use $H_f=m_{\rm loc}h_f$, where $H_f$ is the first-quantized local impurity Hamiltonian matrix of the effective quadratic model, $h_f$ is the dimensionless class-dependent representative matrix listed below, and $m_{\rm loc}=10$. \label{tab:az-local-hf}}
	    \begin{ruledtabular}
	    \begin{tabular}{ccccc}
	        Class & $U_T$ & $U_C$ & $U_S$ & example $h_f$ \\
	        \hline
	        A & -- & -- & -- & $Q_{N_f}$ \\
	        AIII & -- & -- & $\mathbb{1}_{N_f}$ & $0$ \\
	        AI & $\mathbb{1}_{N_f}$ & -- & -- & $M_{N_f}$ \\
	        BDI & $\mathbb{1}_{N_f}$ & $\mathbb{1}_{N_f}$ & $\mathbb{1}_{N_f}$ & $0$ \\
	        D & -- & $\mathbb{1}_{N_f}$ & -- & $\ii\Omega_{N_f}$ \\
	        DIII & $J_{N_f}$ & $\mathbb{1}_{N_f}$ & $J_{N_f}$ & $\ii\Omega_{N_f/2}\otimes\sigma_z$ \\
	        AII & $J_{N_f}$ & -- & -- & $M_{N_f/2}\otimes\sigma_0+\ii\lambda\Omega_{N_f/2}\otimes\sigma_x$ \\
	        CII & $J_{N_f}$ & $J_{N_f}$ & $-\mathbb{1}_{N_f}$ & $0$ \\
	        C & -- & $J_{N_f}$ & -- & $M_{N_f/2}\otimes\sigma_z+\lambda(D_{N_f/2}-X_{N_f/2})\otimes\sigma_y$ \\
	        CI & $\mathbb{1}_{N_f}$ & $J_{N_f}$ & $J_{N_f}$ & $M_{N_f/2}\otimes\sigma_z$
	    \end{tabular}
	    \end{ruledtabular}
	\end{table*}

\begin{figure}[t]
	    \includegraphics[width=\linewidth]{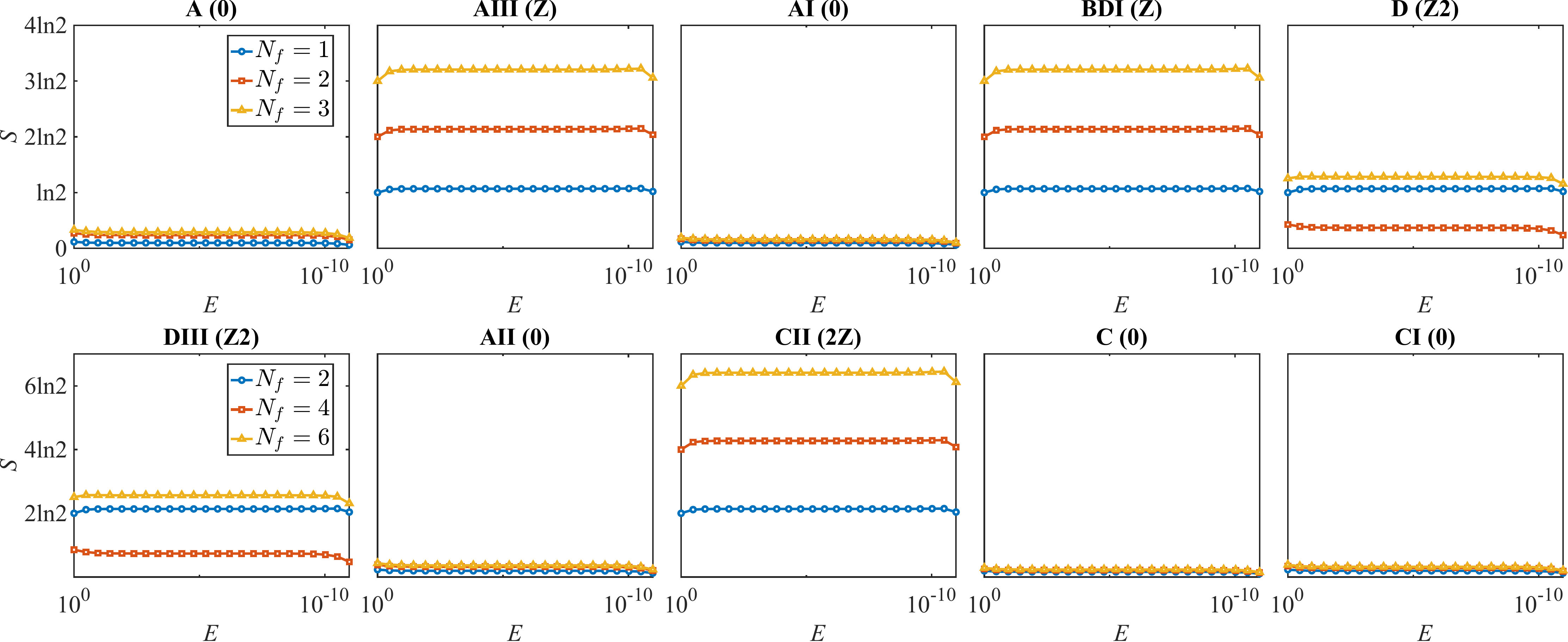}
	    \caption{Entanglement entropy for the ten AZ classes with constant identity hybridization $\Delta(\omega)=\mathbb{1}_{N_f}$ and symmetry-allowed local impurity Hamiltonian $H_f$. The local terms are defined in \cref{tab:az-local-hf,eq:direct-local-building-blocks,eq:local-mass-representative}. The calculation uses bandwidth $D=1$, $\Lambda=3$, and $N_{\rm shell}=24$ logarithmic bath shells. The labels $N_f$ count all single-particle flavors of the impurity/hybridization matrix, including spin and orbital labels; an $N$-orbital spin-$1/2$ system therefore has $N_f=2N$. Topologically trivial classes are pushed close to zero residual entropy by the local mass, while the chiral classes AIII, BDI, and CII are unaffected because $H_f=0$ is symmetry enforced. The D and DIII classes show the expected $\mathbb{Z}_2$ parity effect. \label{fig:az10-localmass-appendix}}
	\end{figure}

For the local-mass calculation in \cref{fig:az10-localmass-appendix}, the following impurity Hamiltonian is one concrete representative example rather than a unique choice. The bath is kept as a constant identity hybridization, $\Delta(\omega)=\mathbb{1}_{N_f}$, and the impurity term is specified directly as follows. For dimension $r$, define
	\begin{align}
	    D_r &= \sum_{a=1}^{r}d_a |a\rangle\langle a|,\nonumber\\
	    X_r &= x\sum_{a=1}^{r-1}
	    \left(|a\rangle\langle a+1|+|a+1\rangle\langle a|\right),\nonumber\\
	    \Omega_r &= y\sum_{a=1}^{r-1}
	    \left(|a\rangle\langle a+1|-|a+1\rangle\langle a|\right),\nonumber\\
	    M_r &=D_r+X_r,\qquad Q_r=M_r+\ii\Omega_r .
	    \label{eq:direct-local-building-blocks}
	\end{align}
	We then take
	\begin{align}
	    H_f=m_{\rm loc}h_f,
	    \label{eq:local-mass-representative}
	\end{align}
	with $h_f$ listed in \cref{tab:az-local-hf}. For the plotted data we take $d_a=a$, $x=1/3$, $y=2/3$, $m_{\rm loc}=10$  and $\lambda=1/2$. This representative preserves the target AZ symmetries while using simple nearest-neighbor matrix elements to lift accidental degeneracies and, whenever the dimension permits, to break symmetries that are absent in the target class. The additional $\lambda$ terms in AII and C remove the most direct accidental spinless time-reversal or CI symmetry for $N_f>2$; in minimal Kramers or Majorana dimensions, some accidental degeneracies are unavoidable. In the AIII, BDI, and CII representatives the constraints force $H_f=0$, so the residual entropy cannot be reduced by a local impurity term alone.

	\begin{figure}[t]
	    \includegraphics[width=\linewidth]{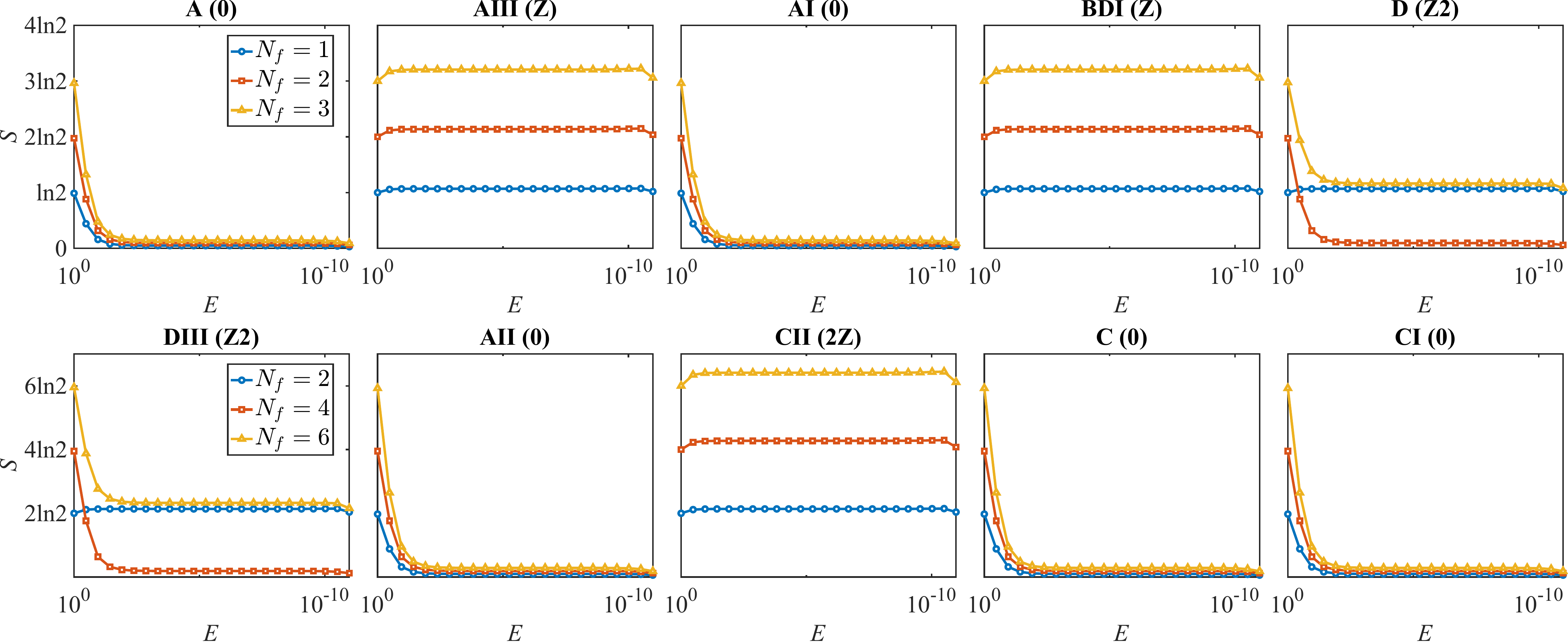}
	    \caption{Entanglement entropy of the effective energy-space Hamiltonian for the ten Altland-Zirnbauer symmetry classes. The calculation uses bandwidth $D=1$, $\Lambda=3$, and $N_{\rm shell}=24$ logarithmic bath shells. For each class we set the impurity single-particle Hamiltonian to zero and use the clipped-linear spectral hybridization construction described in the text: $\Delta_{\rm raw}(\omega)=\Delta^{(0)}+\omega\Delta^{(1)}$, followed by eigenvalue clipping to ensure $\Delta(\omega)\ge0$. The linear scale is $c_{\rm lin}=8$. The labels $N_f$ count all single-particle flavors of the hybridization matrix, including spin and orbital labels; an $N$-orbital spin-$1/2$ system therefore has $N_f=2N$. The first row has minimal dimension one, so the three curves correspond to $N_f=1,2,3$; the second row has minimal dimension two, so the curves correspond to $N_f=2,4,6$. The chiral classes exhibit the $\mathbb{Z}$ or $2\mathbb{Z}$ plateaus, whereas the D and DIII classes exhibit the $\mathbb{Z}_2$ dependence on the parity of the number of units. \label{fig:az10-appendix}}
	\end{figure}

	To make clear that the same construction can also be implemented on the bath side, \cref{fig:az10-appendix} shows one representative clipped-linear hybridization protocol with $H_f=0$ for all AZ classes. This protocol is an example, not a unique way to realize the corresponding symmetry classes. The bandwidth is set to $D=1$, and the Wilson-shell frequencies are $\omega=\pm\varepsilon_i$ with $\varepsilon_i=\varepsilon_1\Lambda^{-(i-1)}$, $i=1,\ldots,N_{\rm shell}$.

In this representative example, the raw hybridization function is defined as $\Delta_{\rm raw}(\omega)=\Delta^{(0)}+\omega\Delta^{(1)}$ and is required to preserve the target symmetries. Since a physical hybridization function must be positive semidefinite, we diagonalize $\Delta_{\rm raw}(\omega)$, set its negative eigenvalues to zero, and rotate it back to the original basis. This clipping preserves the symmetry as long as $\Delta_{\rm raw}(\omega)$ preserves the symmetry.

The explicit representatives used in \cref{fig:az10-appendix} are as follows, with $c_{\rm mix}=0.1$ and $c_{\rm lin}=8$ in the plot. For the first-row classes one explicit set is
\begin{align}
{\rm A}:\quad
&\Delta^{(0)}=
\begin{cases}
\mathbb{1}_1, & N_f=1,\\
\mathbb{1}_{N_f}+c_{\rm mix}Y_{12}, & N_f=2,3,
\end{cases}
\qquad
\Delta^{(1)}=c_{\rm lin}\mathbb{1}_{N_f}, \nonumber\\
{\rm AIII}:\quad
&\Delta^{(0)}=
\begin{cases}
\mathbb{1}_1, & N_f=1,\\
\mathbb{1}_{N_f}+c_{\rm mix}Y_{12}, & N_f=2,3,
\end{cases}
\qquad
\Delta^{(1)}=0, \nonumber\\
{\rm AI}:\quad
&\Delta^{(0)}=
\begin{cases}
\mathbb{1}_1, & N_f=1,\\
\mathbb{1}_{N_f}+c_{\rm mix}X_{12}, & N_f=2,3,
\end{cases}
\qquad
\Delta^{(1)}=c_{\rm lin}\mathbb{1}_{N_f}, \nonumber\\
{\rm BDI}:\quad
&\Delta^{(0)}=
\begin{cases}
\mathbb{1}_1, & N_f=1,\\
\mathbb{1}_{N_f}+c_{\rm mix}X_{12}, & N_f=2,3,
\end{cases}
\qquad
\Delta^{(1)}=0, \nonumber\\
{\rm D}:\quad
&\Delta^{(0)}=
\begin{cases}
\mathbb{1}_1, & N_f=1,\\
\mathbb{1}_{N_f}+c_{\rm mix}X_{12}, & N_f=2,3,
\end{cases}
\qquad
\Delta^{(1)}=
\begin{cases}
0, & N_f=1,\\
c_{\rm lin}Y_{12}, & N_f=2,\\
c_{\rm lin}(Y_{12}+Y_{23})/\sqrt2, & N_f=3.
\end{cases}
\end{align}

where $X_{ab}=|a\rangle\langle b|+|b\rangle\langle a|$ and $Y_{ab}=-\ii|a\rangle\langle b|+\ii|b\rangle\langle a|$ for $a,b=1,\ldots,N_f$.

For the second-row classes, $N_f$ must be even due to $\mathcal{T}^2=-1$ or $\mathcal{C}^2=-1$ symmetry. One explicit set is
\begin{align}
{\rm DIII}:\quad
&\Delta^{(0)}=\mathbb{1}_{N_f}+c_{\rm mix}R_{N_f/2}\otimes\mathbb{1}_2,\qquad
\Delta^{(1)}=
\begin{cases}
0, & N_f=2,\\
c_{\rm lin}P^y_{12}, & N_f=4,\\
c_{\rm lin}(P^y_{12}+P^y_{23})/\sqrt2, & N_f=6,
\end{cases}
\nonumber\\
{\rm AII}:\quad
&\Delta^{(0)}=
\begin{cases}
\mathbb{1}_2, & N_f=2,\\
\mathbb{1}_{N_f}+c_{\rm mix}P^x_{12}, & N_f=4,6,
\end{cases}
\qquad
\Delta^{(1)}=c_{\rm lin}\mathbb{1}_{N_f}, \nonumber\\
{\rm CII}:\quad
&\Delta^{(0)}=
\begin{cases}
\mathbb{1}_2, & N_f=2,\\
\mathbb{1}_{N_f}+c_{\rm mix}P^x_{12}, & N_f=4,6,
\end{cases}
\qquad
\Delta^{(1)}=0, \nonumber\\
{\rm C}:\quad
&\Delta^{(0)}=\mathbb{1}_{N_f}+c_{\rm mix}R_{N_f/2}\otimes\mathbb{1}_2,\qquad
\Delta^{(1)}=c_{\rm lin}(\mathbb{1}_{N_f/2}\otimes\sigma_y), \nonumber\\
{\rm CI}:\quad
&\Delta^{(0)}=\mathbb{1}_{N_f}+c_{\rm mix}R_{N_f/2}\otimes\mathbb{1}_2,\qquad
\Delta^{(1)}=c_{\rm lin}(\mathbb{1}_{N_f/2}\otimes\sigma_z).
\end{align}

where $R_1=0$ and $R_m={\rm diag}(m-1,m-3,\ldots,1-m)/(m-1)$ for $m>1$, while $P^x_{pq}=X_{2p-1,2q-1}+X_{2p,2q}$ and $P^y_{pq}=Y_{2p-1,2q-1}-Y_{2p,2q}$ for $p,q=1,\ldots,N_f/2$.

The low-energy entropy follows the symmetry class and the topology of the corresponding effective energy-space Hamiltonian, not the particular representative chosen above, which we have checked by using other randomly chosen symmetry-allowed $\Delta(\omega)$ in the corresponding class. We find that the chiral classes with $\mathbb{Z}$ or $2\mathbb{Z}$ classification show the expected integer or even-integer $\ln2$ plateaus, the D and DIII classes show the $\mathbb{Z}_2$ parity effect, and the topologically trivial classes have no protected residual entropy apart from nonuniversal corrections.

\begin{figure}[t]
    \centering
    \includegraphics[width=0.4\linewidth]{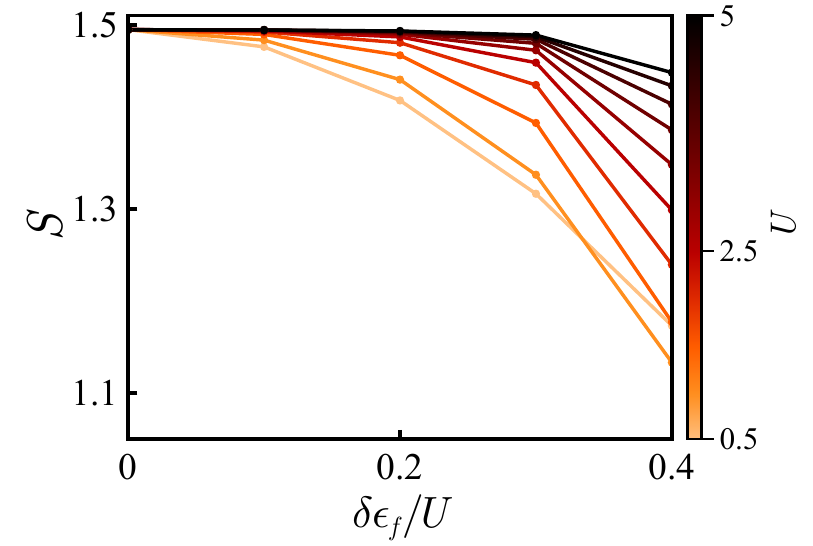}
    \caption{Fixed-point entropy $S$ of the one-orbital Anderson impurity model with constant hybridization as a function of $\delta\epsilon_f/U$, where $\delta\epsilon_f=\epsilon_f+U/2$ measures the deviation of the impurity level from the charge-neutral point. The calculation uses $\Delta_0=0.05$, $D=1$, $\Lambda=2$, and colors denote $U=0.5,1.0,\ldots,5.0$. \label{fig:onsite-asym-appendix}}
\end{figure}

\paragraph{Relation to realistic models.} The above discussion concerns generic impurity fixed points within a given AZ symmetry class. We now explain how these classes can be related to more microscopic impurity models. In a number-conserving Anderson or Kondo problem, chiral symmetry may emerge from a sublattice structure, while particle-hole symmetry may emerge near half filling.

When $\tk\!\ll\!\Delta(0)$, the impurity level is pinned close to zero energy \cite{hewson_renormalized_1993}, so the low-energy fixed point may be viewed as having $H_f\simeq0$. This is consistent with the constant-hybridization scan in \cref{fig:onsite-asym-appendix}, where the fixed-point entropy remains close to the BDI value when the impurity-level asymmetry is small on the low-energy scale. If in addition $\tk\cdot d\Delta/d\omega \!\ll\! \Delta(0)$, the hybridization is approximately constant over the Kondo scale; this bath-side condition is tested in \cref{fig:topo}(c). The energy-space Hamiltonian then has the chiral symmetry with $U^f_S=\mathbb{1}$. When the physical time-reversal symmetry satisfies $\mathcal{T}^2=-1$ and spin $SU(2)$ symmetry is present, one can combine time reversal with a spin rotation to define $\mathcal{T}'=\mathcal{T} e^{\ii \pi S_y}$, which satisfies $(\mathcal{T}')^2=1$ and does not mix the two spin sectors. Thus a generic fully screened Anderson or Kondo fixed point with time reversal and spin rotation symmetry belongs to class BDI within each spin sector. For an $N$-orbital impurity, the two spin sectors give $N_f=2N$ and hence the BDI plateau $S_\Lambda(0)=2N\ln2$, up to the usual $\Lambda$-dependent corrections, as shown in \cref{fig:topo}. In \cref{fig:topo}(d), $N$ labels the number of orbitals in BDI and the number of Kramers-pair units in DIII, with $N_f=2N$ in either panel; \cref{fig:az10-appendix} labels the total flavor number $N_f$ directly.

A magnetic field or spin-orbit coupling changes this conclusion by breaking some of the above emergent symmetries. A magnetic field breaks time reversal, while spin-orbit coupling can preserve the physical $\mathcal{T}^2=-1$ symmetry but breaks the spin rotation used to define $\mathcal{T}'$. If these perturbations are small compared with the Kondo scale, so that $H_f\simeq0$ and the low-energy chiral symmetry remain good approximations, the effective classes are reduced to AIII or CII. If they are large enough to split the impurity level on the scale of the fixed-point physics, the chiral symmetry is also broken and the low-energy class is reduced to A or AII, for which the residual entropy is not topologically protected.

Superconducting impurity problems provide another route to the D and DIII classes, but the particle-hole operation should then be interpreted differently. In a BdG description it is the intrinsic redundancy of the Nambu or Majorana representation, rather than an additional charge-conjugation symmetry of a number-conserving Hamiltonian. A fully gapped bulk superconductor does not by itself provide the constant low-energy hybridization assumed above; such a description is instead appropriate for gapless superconducting sectors with finite local spectral weight, for example chiral or helical Majorana boundary modes, or nodal quasiparticles in a suitable energy window. In class D, a closed BdG Hamiltonian built from ordinary complex electron modes has an even-dimensional Nambu space. Therefore, odd-dimensional class-D representatives should not be regarded as ordinary finite-dimensional BdG Hamiltonians. They are better viewed as Majorana low-energy theories, such as an odd number of chiral Majorana edge channels coupled to an impurity sector containing an odd number of local Majorana zero modes. In class DIII, a microscopic electronic BdG Hamiltonian with physical time reversal has a Nambu dimension that is a multiple of four. If the effective dimension is not a multiple of four, the impurity sector should be understood as containing Majorana Kramers pairs. For a pair $\gamma_1,\gamma_2$ one may choose $\mathcal{T}\gamma_1\mathcal{T}^{-1}=\gamma_2$ and $\mathcal{T}\gamma_2\mathcal{T}^{-1}=-\gamma_1$. Although they can be combined into a complex fermion $f=(\gamma_1+\ii\gamma_2)/2$, time reversal maps this fermion to its hole, $\mathcal{T}f\mathcal{T}^{-1}=\ii f^\dagger$. Thus the minimal D and DIII representatives should be understood as effective Majorana boundary/impurity theories, while a fully microscopic electronic BdG realization requires embedding them into a larger Nambu space with the BdG double counting properly removed.

\section{Details of DMRG}
\subsection{Matrix product operator representation of the Hamiltonian \label{app:MPO}}
In the DMRG calculation, the Hamiltonian is represented as a matrix product operator (MPO). We use the ITensor Julia libraries \cite{ITensor,ITensor-r0.3,corbett_scaling_2025}, which automatically construct the MPO from an operator sum. Although the star representation is long-ranged in the one-dimensional ordering used by the MPS, the long-range part has a separable structure in the bath indices. Therefore the required MPO bond dimension does not grow with the number of logarithmic bath shells. Such finite-bond-dimension forms can be obtained from the standard finite-automaton construction of MPOs \cite{crosswhite_finite_2008}. The formulas below give explicit algebraic MPO examples showing this structure; they are not unique and need not be the optimal MPO used in practice. For compactness, the examples may group a fixed number of internal components into composite sites; resolving these components changes the MPO bond dimension only by an $L$-independent factor. The entries are written as global fermionic operators obeying the canonical anticommutation relations, so the displayed matrix products can be expanded literally. A local tensor-network implementation can realize the same expressions either as a $\mathbb{Z}_2$-graded MPO or as an ordinary MPO after inserting the corresponding Jordan-Wigner parity strings; these implementation are not written explicitly here.

For the Anderson impurity model, we start from the discretized Hamiltonian in \cref{eq:HAIM-dis},
\begin{align}
    \hH_{\rm AIM}
    =&\ \hH_f+\sum_{\alpha s}\sum_{i=1}^{L}\sum_{\eta=\pm}
    \eta\varepsilon_i c^\dagger_{i\eta\alpha s}c_{i\eta\alpha s} +\sum_{\alpha s}\sum_{i=1}^{L}\sum_{\eta=\pm}
    \left(t_{i\eta}c^\dagger_{i\eta\alpha s}f_{\alpha s}
    +t_{i\eta}^*f^\dagger_{\alpha s}c_{i\eta\alpha s}\right),
    \label{eq:dmrg-aim-star}
\end{align}
where $|t_{i\eta}|^2=\int_{i,\eta}{\rm d}\varepsilon\,\Delta(\varepsilon)$ as in \cref{app:discretization}. The site label in the MPS is the compound energy label $(i,\eta)$, together with its spin and orbital flavor. For a fixed spin-orbital flavor $(\alpha,s)$, define
\begin{align}
    \hH_{\alpha s}^{\rm AIM}
    =&\ \sum_{i=1}^{L}\sum_{\eta=\pm}
    \eta\varepsilon_i \hat n_{i\eta\alpha s}
    +\sum_{i=1}^{L}\sum_{\eta=\pm}
    \left(t_{i\eta}c^\dagger_{i\eta\alpha s}f_{\alpha s}
    +t_{i\eta}^*f^\dagger_{\alpha s}c_{i\eta\alpha s}\right),
    \label{eq:dmrg-aim-channel}
\end{align}
so that $\hH_{\rm AIM}=\hH_f+\sum_{\alpha s}\hH_{\alpha s}^{\rm AIM}$. The important point for the MPO construction is that the impurity-bath coupling in each $\hH_{\alpha s}^{\rm AIM}$ contains only one bath operator. The minus sign in the impurity tensor below is needed because the literal expansion gives $-f_{\alpha s}c^\dagger_{\ell\alpha s}=c^\dagger_{\ell\alpha s}f_{\alpha s}$. Let $\ell=1,\ldots,2L$ enumerate the bath sites in the MPS order, with $\ell\leftrightarrow(i_\ell,\eta_\ell)$, $e_\ell=\eta_\ell\varepsilon_{i_\ell}$, and $t_\ell=t_{i_\ell\eta_\ell}$. One explicit open-boundary MPO for $\hH_{\alpha s}^{\rm AIM}$ is
\begin{align}
W_{f,\alpha s}^{\rm AIM}
=\begin{pmatrix}
0 & -f_{\alpha s} & f^\dagger_{\alpha s} & I
\end{pmatrix}.
\label{eq:dmrg-aim-first-mpo}
\end{align}
For all bath sites before the terminal one, the tensor may be chosen as
\begin{align}
W_{\ell,\alpha s}^{\rm AIM}=
\begin{pmatrix}
I & 0 & 0 & 0\\
t_{\ell}c^\dagger_{\ell\alpha s} & I & 0 & 0\\
t_{\ell}^*c_{\ell\alpha s} & 0 & I & 0\\
e_\ell \hat n_{\ell\alpha s} & 0 & 0 & I
\end{pmatrix},
\label{eq:dmrg-aim-mpo}
\end{align}
whereas the terminal bath tensor is the closing column
\begin{align}
W_{2L,\alpha s}^{\rm AIM}
=
\begin{pmatrix}
I\\
t_{2L}c^\dagger_{2L,\alpha s}\\
t_{2L}^*c_{2L,\alpha s}\\
e_{2L} \hat n_{2L,\alpha s}
\end{pmatrix}.
\label{eq:dmrg-aim-terminal-mpo}
\end{align}
The product
\begin{align}
W_{f,\alpha s}^{\rm AIM}
\left(\prod_{\ell=1}^{2L-1} W_{\ell,\alpha s}^{\rm AIM}\right)
W_{2L,\alpha s}^{\rm AIM}
\end{align}
then gives $\hH_{\alpha s}^{\rm AIM}$. The remaining impurity term $\hH_f$ also has finite bond dimension. For an $N$-orbital Anderson model, there are $2N$ spin-orbital flavors, so the full Hamiltonian is the sum of $2N+1$ finite-bond-dimension MPOs: one for $\hH_f$ and one for each $\hH_{\alpha s}^{\rm AIM}$. This sum can be written as a single MPO by taking the direct sum of the corresponding auxiliary spaces, and the resulting bond dimension is still finite and independent of $L$.

For the special case in which the discretized coefficients are strictly exponential,
\begin{align}
t_{i\eta}=t_{\eta}\Lambda^{-(i-1)/2},
\qquad
\varepsilon_i=\varepsilon_1\Lambda^{-(i-1)},
\label{eq:dmrg-aim-exponential}
\end{align}
the same channel MPO can be written in a translation-invariant form. The first tensor is still $W_{f,\alpha s}^{\rm AIM}$ in \cref{eq:dmrg-aim-first-mpo}. Since the MPS ordering used here is $(-\varepsilon_i,+\varepsilon_i)$ within each shell, one convenient choice is to put the scale factors on the identity lines of the second site in each two-site shell. Apart from the terminal site, the two bath tensors are then independent of the shell index $i$:
\begin{align}
\widetilde W_{i,-,\alpha s}^{\rm AIM}
&=
\begin{pmatrix}
I & 0 & 0 & 0\\
t_{-}c^\dagger_{i,-,\alpha s} & I & 0 & 0\\
t_{-}^*c_{i,-,\alpha s} & 0 & I & 0\\
-\varepsilon_1 \hat n_{i,-,\alpha s} & 0 & 0 & I
\end{pmatrix},
\nonumber\\
\widetilde W_{i,+,\alpha s}^{\rm AIM}
&=
\begin{pmatrix}
I & 0 & 0 & 0\\
t_{+}c^\dagger_{i,+,\alpha s} & \Lambda^{-1/2}I & 0 & 0\\
t_{+}^*c_{i,+,\alpha s} & 0 & \Lambda^{-1/2}I & 0\\
\varepsilon_1 \hat n_{i,+,\alpha s} & 0 & 0 & \Lambda^{-1}I
\end{pmatrix}.
\label{eq:dmrg-aim-scale-invariant-mpo}
\end{align}
Multiplying the tensors through the first $i-1$ shells accumulates the factors $\Lambda^{-(i-1)/2}$ for the hybridization terms and $\Lambda^{-(i-1)}$ for the onsite-energy terms, thereby reproducing \cref{eq:dmrg-aim-exponential}.

For the Kondo model, the discretized Hamiltonian in \cref{eq:HK-dis} is
\begin{align}
    \hH_K
    =&\ \sum_{\alpha s}\sum_{i,\eta}
    \eta\varepsilon_i c^\dagger_{i\eta\alpha s}c_{i\eta\alpha s}
    +\frac{J_K}{2}\sum_{\alpha ss'}
    \hat{\mathbf S}_f\cdot{\bm\sigma}_{ss'} \times\sum_{i,i'=1}^{L}\sum_{\eta,\eta'=\pm}
    t_{i\eta}t_{i'\eta'}
    c^\dagger_{i\eta\alpha s}c_{i'\eta'\alpha s'} .
    \label{eq:dmrg-kondo-star}
\end{align}
For each channel $\alpha$, let $\ell=1,\ldots,2L$ enumerate the bath energy sites, with $\ell\leftrightarrow(i_\ell,\eta_\ell)$, and define $e_\ell=\eta_\ell\varepsilon_{i_\ell}$ and $t_\ell=t_{i_\ell\eta_\ell}$, where $t_{i,\pm}^2=\int_{i,\pm}{\rm d}\varepsilon\,\nu(\varepsilon)$ and we choose $t_{i,\pm}>0$. The interaction part can be written explicitly as
\begin{align}
    \hH_{K,{\rm int}}
    =&\ \sum_{\alpha}\sum_{\ell\ell'} t_{\ell}t_{\ell'}\Big[
    \frac{J_K}{2}S_f^+
    c^\dagger_{\ell\alpha\down}c_{\ell'\alpha\up}
    +\frac{J_K}{2}S_f^-
    c^\dagger_{\ell\alpha\up}c_{\ell'\alpha\down} +\frac{J_K}{2}S_f^z
    c^\dagger_{\ell\alpha\up}c_{\ell'\alpha\up}
    -\frac{J_K}{2}S_f^z
    c^\dagger_{\ell\alpha\down}c_{\ell'\alpha\down}
    \Big].
    \label{eq:dmrg-kondo-spin-resolved}
\end{align}
This spin-resolved form makes the compact MPO possible because every two-bath-site coefficient factorizes as $t_\ell t_{\ell'}$. For compactness, the algebraic MPO below groups the two spin components at fixed $(\ell,\alpha)$ into one effective spinful bath site. One explicit MPO for a fixed bath channel $\alpha$ is
\begin{align}
    W_{f,\alpha}^{K}
    =\begin{pmatrix}
    0 & \mathbf{S}_f & \mathbf{S}_f & \mathbf{0} & \mathbf{0} & I
    \end{pmatrix},
    \qquad
    \mathbf{S}_f=(S_f^+,S_f^-,S_f^z,S_f^z),
    \label{eq:dmrg-kondo-imp-mpo}
\end{align}
on the impurity site and
\begin{align}
W_{\ell,\alpha}^{K}
=
\begin{pmatrix}
I & \mathbf{0} & \mathbf{0} & \mathbf{0} & \mathbf{0} & 0\\
\mathbf{L}_{\ell\alpha} & I_4 & 0 & \mathbf{A}^{>}_{\ell\alpha} & 0 & \mathbf{0}\\
0 & 0 & I_4 & 0 & \mathbf{A}^{<}_{\ell\alpha} & \mathbf{0}\\
\mathbf{B}^{>}_{\ell\alpha} & 0 & 0 & I_4 & 0 & \mathbf{0}\\
\mathbf{B}^{<}_{\ell\alpha} & 0 & 0 & 0 & I_4 & \mathbf{0}\\
h_{\ell\alpha} & \mathbf{0} & \mathbf{0} & \mathbf{0} & \mathbf{0} & I
\end{pmatrix},
\label{eq:dmrg-kondo-bath-mpo}
\end{align}
on all bath sites before the terminal one. Here $I_4$ is the identity on the four-component auxiliary space, and bold zeros denote zero blocks with the dimensions implied by the neighboring entries. The open-boundary MPO for this fixed bath channel is
\begin{align}
W_{f,\alpha}^{K}
\left(\prod_{\ell=1}^{2L-1}W_{\ell,\alpha}^{K}\right)
W_{2L,\alpha}^{K},
\end{align}
where the terminal bath-site tensor is
\begin{align}
W_{2L,\alpha}^{K}
=
\begin{pmatrix}
I\\
\mathbf{L}_{2L,\alpha}\\
\mathbf{0}\\
\mathbf{B}^{>}_{2L,\alpha}\\
\mathbf{B}^{<}_{2L,\alpha}\\
h_{2L,\alpha}
\end{pmatrix}.
\label{eq:dmrg-kondo-terminal-mpo}
\end{align}
Here
\begin{align}
    h_{\ell\alpha}&=e_\ell(\hat n_{\ell\alpha\up}+\hat n_{\ell\alpha\down}),\\
    \mathbf{A}^{>}_{\ell\alpha}
    &=t_\ell{\rm diag}
    (c^\dagger_{\ell\alpha\down},
    c^\dagger_{\ell\alpha\up},
    c^\dagger_{\ell\alpha\up},
    c^\dagger_{\ell\alpha\down}),\\
    \mathbf{B}^{>}_{\ell\alpha}
    &=\frac{J_K t_\ell}{2}
    \begin{pmatrix}
    c_{\ell\alpha\up}\\
    c_{\ell\alpha\down}\\
    c_{\ell\alpha\up}\\
    -c_{\ell\alpha\down}
    \end{pmatrix},
    \qquad
    \mathbf{L}_{\ell\alpha}
    =\frac{J_K t_\ell^2}{2}
    \begin{pmatrix}
    c^\dagger_{\ell\alpha\down}c_{\ell\alpha\up}\\
    c^\dagger_{\ell\alpha\up}c_{\ell\alpha\down}\\
    \hat n_{\ell\alpha\up}\\
    -\hat n_{\ell\alpha\down}
    \end{pmatrix},\\
    \mathbf{A}^{<}_{\ell\alpha}
    &=t_\ell{\rm diag}
    (c_{\ell\alpha\up},
    c_{\ell\alpha\down},
    c_{\ell\alpha\up},
    c_{\ell\alpha\down}),\\
    \mathbf{B}^{<}_{\ell\alpha}
    &=\frac{J_K t_\ell}{2}
    \begin{pmatrix}
    -c^\dagger_{\ell\alpha\down}\\
    -c^\dagger_{\ell\alpha\up}\\
    -c^\dagger_{\ell\alpha\up}\\
    c^\dagger_{\ell\alpha\down}
    \end{pmatrix}.
    \label{eq:dmrg-kondo-blocks}
\end{align}
Expanding the product, the blocks with superscript $>$ give the terms in which $c^\dagger_\ell$ appears before $c_{\ell'}$ in the MPS ordering, while the blocks with superscript $<$ give the fermion-reordered partner terms. The vector $\mathbf{L}_{\ell\alpha}$ gives the $\ell=\ell'$ contribution. For several bath channels, the same fixed-$\alpha$ construction is repeated in a larger block matrix and summed over $\alpha$. Therefore the maximum MPO bond dimension is fixed by the number of channels and spin components, while increasing $L$ only repeats the same bath tensor more times.

\subsection{Numerical setup}
\paragraph{Sites and symmetries.} We order the sites according to the star geometry used in the main text. The impurity degrees of freedom are placed at the left end of the MPS chain, followed by bath modes ordered from high to low energy. The positive- and negative-energy modes in the same logarithmic shell and their internal components are kept adjacent. In the Anderson models, every orbital--spin component of both the impurity and the bath is represented by a separate fermionic MPS site. For the single-orbital model we exploit the total U(1) particle-number symmetry. For the two-orbital model we separately conserve the particle number in each orbital and the total $S^z$. In the Kondo models, the impurity spin is represented by a single spin-$S_{\rm imp}$ site, while every channel--spin component of a bath mode is represented by a separate fermionic site. We conserve the particle number in each screening channel together with the total $S^z$. The entanglement cut therefore separates the impurity and the bath modes above $\varepsilon_i$ from the remaining lower-energy bath modes.

\paragraph{Initialization.} For the Anderson models, the initial MPS is randomized within the symmetry sector selected by a half-filled reference configuration. For the Kondo models, we start from a product MPS in which the negative-energy bath modes are occupied and the impurity is placed in the appropriate spin sector. All calculations are performed at half filling.

\paragraph{Details of the sweeps.} We use two-site DMRG. Most of the data shown in the main text use a maximum MPS bond dimension $\chi=1600$; the spin-$\frac12$ three-channel Kondo scan in \cref{fig:NFL}(d) is restarted from $\chi=6400$ and continued to $\chi=12800$. The maximum truncation error reported in the final sweep is approximately $10^{-12}$--$10^{-9}$ for the one-orbital Anderson data and is of order $10^{-14}$ for the one-channel Kondo data. For the two-channel Kondo calculations it ranges from $3\times10^{-11}$ to $9\times10^{-9}$. The more entangled two-orbital Anderson calculations reach $\chi=1600$ over an extended part of the chain; their final-sweep maximum truncation errors range from $4\times10^{-10}$ to $6\times10^{-6}$. The three-channel Kondo calculations reach $\chi=12800$, with final-sweep maximum truncation errors from $6\times10^{-10}$ to $1.3\times10^{-8}$.

\begin{figure*}[t]
    \includegraphics[width=0.7\textwidth]{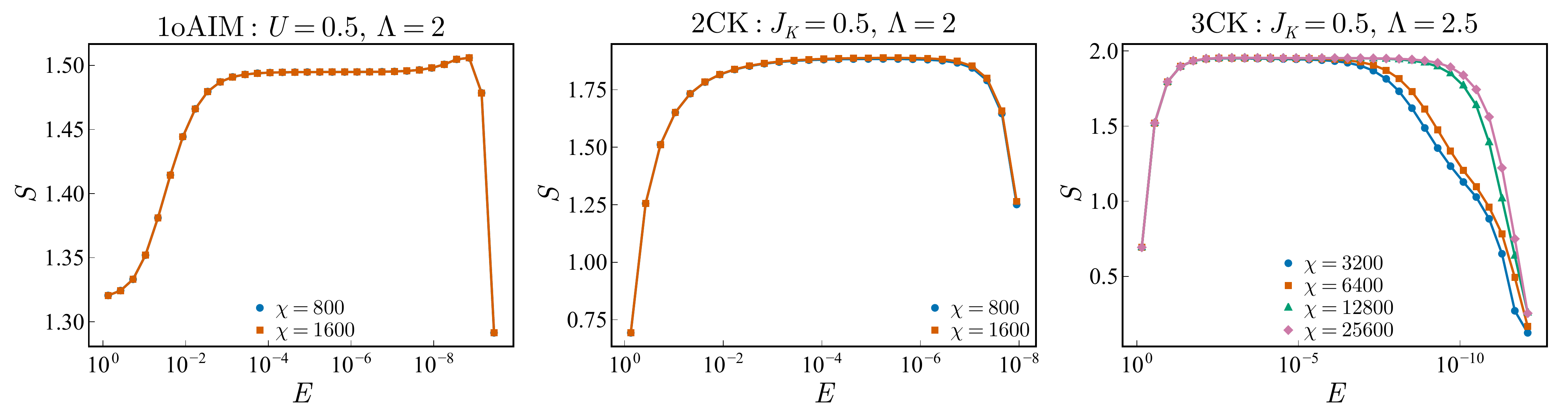}
    \caption{Bond-dimension dependence of the energy-space entanglement entropy $S_\Lambda(E)$ for representative calculations. (a) One-orbital Anderson impurity model with $U=0.5$ and $\Lambda=2$, comparing $\chi=800$ and $1600$. (b) Two-channel Kondo model with $J_K=0.5$ and $\Lambda=2$, comparing $\chi=800$ and $1600$. (c) Three-channel Kondo model with $J_K=0.5$ and $\Lambda=2.5$, comparing $\chi=3200$, $6400$, $12800$, and $25600$. The one-orbital Anderson curves are indistinguishable on the scale of the plot. In the multichannel Kondo models, the remaining bond-dimension dependence is concentrated near the lowest-energy crossover, while the fixed-point plateau value is stable at the largest bond dimensions used. \label{fig:bond-dimension-convergence}}
\end{figure*}

Direct comparisons with $\chi=800$ show that the one-orbital Anderson $U$ and $\Lambda$ scans and the one-channel Kondo $\Lambda$ scan are already converged at the smaller bond dimension: over all corresponding parameter points, reducing $\chi$ from 1600 to 800 changes the displayed entanglement profiles by at most $8.2\times10^{-6}$ and the extracted fixed-point entropies by at most $6.7\times10^{-6}$. Thus, for these observables, the truncation error need not be reduced to the smallest values reached in the $\chi=1600$ runs; for example, the one-orbital Anderson $\Lambda$ scan remains converged when the final-sweep maximum truncation error at $\chi=800$ is as large as $6.9\times10^{-9}$. The two-channel Kondo scan is not uniformly converged at $\chi=800$, and the three-channel Kondo scan is not converged at $\chi=6400$, so we retain the larger bond dimensions for those data. As shown in \cref{fig:bond-dimension-convergence}, direct bond-dimension comparisons confirm that the plateau of $S_\Lambda(E)$ is nearly converged: although the lowest-energy crossover in the multichannel Kondo models retains some $\chi$ dependence, the bond dimensions used to extract the fixed-point entropy are sufficient to determine the plateau value. The local eigensolver uses a Krylov-subspace dimension of approximately $10$--$30$. Empirically, calculations at larger $\Lambda$ require a Krylov dimension near the upper end of this range for stable convergence. The sweeps begin with a relatively large density-matrix noise to escape metastable states; the noise is then reduced successively over several orders of magnitude and is set to zero during the final convergence sweeps. The calculations are continued until the energy and the entanglement profile are stable under further sweeps.

\end{document}